\documentclass[pra,twocolumn]{revtex4}
\usepackage{graphicx,color}
\begin{document}

\title{Many Body Approach for Quartet Condensation in Strong Coupling.}   

\author{Takaaki Sogo, Gerd R\"opke}
\affiliation{Institut f\"ur Physik, Universit\"at Rostock, 
             D-18051 Rostock, Germany}
\author{Peter Schuck}
\affiliation{
Institut de Physique Nucl\'eaire, CNRS, UMR 8608, Orsay F-91406, France}
\affiliation{Universit\'e Paris-Sud, Orsay F-91505, France}
\affiliation{
Laboratoire de Physique et Mod\'elisation 
des Milieux Condens\'es, CNRS and Universit\'e Joseph Fourier, 
25 Avenue des Martyrs, Bo\^ite Postale 166, F-38042 Grenoble Cedex 9, France}
\affiliation{Groupe de Physique Theorique Institut de Physique Nucleaire,
91406 Orsay Cedex, France}

\begin{abstract}

The theory for condensation of higher fermionic clusters is developed.
Fully selfconsistent nonlinear equations for the quartet 
order parameter 
in strongly coupled fermionic systems
are established and solved.
The breakdown of the quasiparticle picture is pointed out.
Derivation of numerically tractable approximation is described.
The momentum projected factorisation ansatz 
of Ref.~\cite{slr09} for the order parameter is employed again.
As a definite example the condensation of $\alpha$ particles in nuclear 
matter is worked out.
\end{abstract}

\pacs{21.65.-f,67.85.Lm,74.20.Fg}

\maketitle

\section{Introduction}

Quartet condensation is relevant in several domains of physics. It is 
so far mostly considered in nuclear physics with the 
strongly bound alpha particle cluster playing a dominant role in certain 
states of lighter nuclei~\cite{alphaBECfinite} 
and, eventually, also in the surface of heavy nuclei 
as may be indicated by observed alpha decay processes. Alpha particle 
condensation may eventually occur in compact stars~\cite{trk10}. 
However, trapping of 
multi component fermionic atoms makes fast progress. The 
case of trions is already quite advanced, experimentally \cite{whh09,wlo09} 
and also with theory 
\cite{rzh07,gbl08,kdd09,fsw09,edo09,silva09,mds09,acl09,mis10}. 
One may trap 
fermions with four different `colors' in the near future. This is a 
prerequisite for quartet formation and quartet condensation. Theoretical 
work on this subject already has appeared 
\cite{km09,sg99,schlottmann94,wu05,rcl09}. The condensation of bi-
excitons in semi-conductors also may be of relevance~\cite{hn,ms}.

In the recent past we already have published several papers on alpha particle 
condensation in infinite nuclear matter. For instance, we studied the onset of 
alpha particle condensation and evaluated the corresponding critical 
temperature \cite{rss98,slr09} 
with a procedure analogous to the pairing case via a 
four body generalization of the famous Thouless criterion \cite{thouless}. 
However, quartet condensation 
not only is formally much more complicated than 
condensation of (Cooper) pairs, it turns out that also certain aspects of the 
physics are  quite different. The most striking feature being that 
quartets
only exist in the so-called BEC (Bose-Einstein condensate) limit where 
they do not overlap very much in space. 
Contrary to the pairing case quartets cannot strongly intermingle in real 
space and, therefore, a coherence length much longer than the inter alpha 
particle distance cannot exist. We will give the reason for this different 
behavior in the main text.

In this paper we will treat quartet 
condensation at zero temperature and establish and solve a full nonlinear 
equation for the quartet order parameter which will be the analog to the gap 
equation for pairing. A direct solution of such a highly non linear four body 
problem seems hopeless, even in homogeneous matter. However, we recently 
showed that a very simplifying approximation works very well, at least in the 
domain of negative chemical potentials, i.e. in the strong coupling regime. 
This approximation consists in making a mean field ansatz, 
i.e. a Slater determinant, for the quartet but 
projected onto zero total momentum 
as it is relevant for condensation~\cite{slr09}. Such a mean field treatment 
may work each time the quartet is in its lowest energy configuration. 
However, one may think of generalisations for excited configurations as well.
Since the four fermions of the quartet are all different in the 
nuclear case (proton/neutron, spin up/down), 
all are in the same 0S mean field wave 
function and the problem boils down in the end to solve the 
equation in iterating 
on this single one particle mean field wave function. The same happens, 
of course, if the four fermions are components of an $F=3/2$ spin as in [16]. 
The problem still is 
complicated but can be solved with effective interactions of separable form.
It was shown that that approximation gives comparable results as 
with the four body Faddeev-Yakubovsky equation with a more realistic
interaction in the nuclear case. So, in the present work 
we calculate the order parameter equation 
with the projected mean field ansatz and a separable potential.

The paper is organized as follows:
in the next section, we discuss the BCS gap equation.
In Sec.~\ref{sec-quartet}, we show the full expression of 
the single particle mass operator for quartet condensation.
Since this is too difficult to calculate numerically,
we suggest an approximate mass operator in Sec.~\ref{sec-approximate}.
Before we show the numerical result employed by 
the approximated mass operator, 
we discuss in Sec.~\ref{sec-leveldensity} the significant difference between 
pairing condensation and 
quartet condensation through the different level densities involved in the 
condensation processes.
In Sec.~\ref{sec-result},
we present the results.
Finally we conclude in the last section. 
In the Appendix, we describe the detailed 
derivation of the equations and discuss various methods to formulate the order
parameter equation for quartet condensation.

\section{Recapitulation of the pairing case}

In order to prepare the terrain for our procedure in the quartetting case, 
in this section we want to rederive standard BCS theory in a way somewhat 
differing from the usual. 

The one body Green's function (GF) for BCS is represented by 
\cite{rs}

\begin{eqnarray}
G_{1;1'}(\omega)
=\frac{\delta_{11'}}{\omega-\varepsilon_1-M^{\rm BCS}_{1;1}(\omega)},
\label{eq-bcs1pGF}
\end{eqnarray}
 
where $M^{\rm BCS}_{1;1}(\omega)$ is the BCS mass operator

\begin{eqnarray}
M^{\rm BCS}_{1;1'}(\omega)
=
\sum_{2}
\frac{\Delta_{12}\Delta_{1'2}^*}
{\omega+\varepsilon_2},
\label{eq-bcsmass}
\end{eqnarray}
where 
\begin{eqnarray}
\Delta_{12}=-\frac{1}{2}\sum_{34}\bar v_{12,34}\langle c_{4}c_{3}\rangle
\label{eq-pairgap}
\end{eqnarray}
with $\langle cc \rangle$ being the thermal average of the pair operator.

The indices $1,2,3,...$ correspond to momentum and spin. 
In nuclear matter, we have to add isospin.
$\bar v_{12,34}$ is the antisymmetrized matrix 
element of the 
two-body interaction ($\bar v_{12,34}=-\bar v_{12,43}=-\bar v_{21,34}$).
The single particle energies $\varepsilon_i$ are in 
principle given by the kinetic energies plus the mean field shifts. The 
direct term is in homogeneous matter a constant which can be incorporated 
into the chemical potential and the Fock-term gives rise to an effective 
mass. Since we will mostly deal with very low density nuclear matter, we 
will not consider a mass renormalization here. 
Therefore, in Eq.~(\ref{eq-bcs1pGF}) we 
have $\varepsilon_1=k_1^2/(2m)-\mu_1$ with the chemical potential $\mu_1$ 
which, in principle, contains the direct part of the mean field. We have 
attached an index on the chemical potential, since in principal it can depend 
on the various fermionic components involved. However, in this work we always 
will consider fully symmetric situations and henceforth we will suppose that 
the chemical potentials of all fermionic species are equal and, therefore, 
drop the index.

\begin{figure}
\includegraphics[width=75mm]{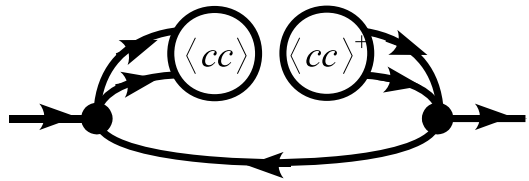}
\caption{\label{fig-bcsmass}Graphic representation of the 
BCS mass operator in Eq.~(\ref{eq-bcsmass})}
\end{figure}

Fig.~\ref{fig-bcsmass} is the graphical representation of the 
BCS mass operator. As shown in Eq.~(\ref{eq-bcsmass}) and 
Figure \ref{fig-bcsmass},
the BCS mass operator consists of 
the 2 particle-1 hole (2p1h) GF between two-body vertices
factorized into two order parameters and the free 1h GF.

On the other hand, the in-medium Schr\"odinger equation 
for the order parameter is of the following form
\begin{eqnarray}
\langle c_{2} c_{1}\rangle
=-\frac{1- \rho_1 - \rho_2}{\varepsilon_1+\varepsilon_2}
\sum_{1'2'}\frac{1}{2}\bar v_{12,1'2'} \langle c_{2'} c_{1'}\rangle,
\label{eq-gapeq}
\end{eqnarray}
where  $\rho_1$ is the occupation number derived from 

\begin{eqnarray}
\rho_{1}
=-\int \frac{d\omega}{2\pi}2{\rm Im} G_{1;1}(\omega+i\eta)
f(\omega)
\label{eq-occupation}
\end{eqnarray}

\noindent
with the Fermi distribution function $f(\omega)=[e^{\omega/T}+1]^{-1}$
and a positive infinitesimal of $\eta$, as indicated.

In the standard BCS theory,
pairs in time reversed states are considered,
i.e. in Eqs.~(\ref{eq-pairgap}) and (\ref{eq-gapeq})
taking $2=\bar 1$, $\rho_{1}=\rho_{\bar 1}$, and
$\varepsilon_{1}=\varepsilon_{\bar 1}$, where $\bar 1$ is
the time reversal conjugate of quantum numbers $1$.
For Eq.~(\ref{eq-bcs1pGF}),
we obtain the imaginary part of the one body Green function as
\begin{eqnarray}
-{\rm Im}G_{1;1}(\omega+i\eta)
&=&
\frac{1}{2}\Bigl(1+\frac{\varepsilon_1}{E_1}\Bigr)
\pi\delta(\omega-E_1)
\nonumber \\
&+&
\frac{1}{2}\Bigl(1-\frac{\varepsilon_1}{E_1}\Bigr)
\pi\delta(\omega+E_1)
\label{eq-imG1BCS}
\end{eqnarray}
with $\Delta_{1\bar 1}=\Delta_{1}$ and 
$E_1=\sqrt{\varepsilon_1^2+\Delta_1^2}$.
This is equivalent to solving the usual gap equation 
at finite temperature as can easily be deduced 
from the spectral function obtained from (\ref{eq-bcs1pGF})~\cite{agd,fw}:
\begin{eqnarray}
\Delta_{1}
=
-\sum_{1'}\frac{1}{2}\bar v_{1 \bar 1,1' \bar 1'} 
\frac{\Delta_{1'}}{2E_{1'}}\tanh(\frac{E_{1'}}{2T}).
\end{eqnarray}

Note that Eq.~(\ref{eq-gapeq}) resembles a particle-particle RPA 
equation \cite{rs} 
with, however, renormalized occupation numbers. One could, therefore, also 
consider Eq.~(\ref{eq-gapeq}) as a single pole approximation to the so-called 
renormalized RPA, well known in the literature, 
see, e.g., Ref.~\cite{hmd02}.

\section{\label{sec-quartet}
Single particle mass operator and quartet condensation}

\begin{figure}
\includegraphics[width=70mm]{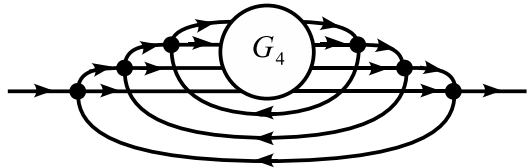}
\caption{\label{fig-diagram}
Some diagram for the mass operator.}
\end{figure}

Formally it is straightforward to generalize the pairing case 
to the quartet case. A typical diagram of the mass operator needed
for quartet condensation is shown in Fig.~\ref{fig-diagram}.
As seen, to make the quartet order parameter appear in the single particle 
mass operator,
we need to express it by the  4p3h GF.
In order to get to it, we must successively open phase space from one particle 
to 2p1h, to 3p2h, to 4p3h. This needs three interaction vertices 
on each side as shown in Fig.~\ref{fig-diagram}. 
Details of the derivation will be given 
in the Appendix~\ref{sec-appendix-alphamass}.  
We obtain for the 1p mass operator with quartet condensation

\begin{eqnarray}
&&
M^{\rm quartet}_{1;1'}(\omega)
\nonumber \\
&=&
\Gamma^{(4)}_{1234;5678}(\omega)
\langle c_{8}c_{7}c_{6}c_{5}\rangle
\frac{\bar f_{2}\bar f_{3}\bar f_{4}+f_{2}f_{3}f_{4}}
     {\omega+\varepsilon_{234}}
     P_{234;2'3'4'}
\nonumber \\
&&\times
\langle c_{5'}^\dagger c_{6'}^\dagger 
              c_{7'}^\dagger c_{8'}^\dagger \rangle
\Gamma^{(4)*}_{1'2'3'4';5'6'7'8'}(\omega),
\label{eq-fullalphaBECmass}
\end{eqnarray}
where summation convention over repeated indices is to be understood,
and
\begin{eqnarray}
P_{123;1'2'3'}
=
\left|
\begin{array}{ccc}
\delta_{11'} & \delta_{12'} & \delta_{13'} \\ 
\delta_{21'} & \delta_{22'} & \delta_{23'} \\ 
\delta_{31'} & \delta_{32'} & \delta_{33'}  
\end{array}
\right|.
\label{eq-antisymmetric}
\end{eqnarray}
 The effective four body vertex $\Gamma^{(4)}$ , evaluated perturbatively to 
third order in the interaction is

\begin{eqnarray}
&&
\Gamma^{(4)}_{1234;5678}(\omega)
\nonumber \\
&=&
\Gamma^{(3)}_{12'3';4'5'6'}(\omega)
\frac{1}{\omega-\varepsilon_{4'5'6'}+\varepsilon_{2'3'}}
\nonumber \\
&\times&
\biggl[
 \frac{1}{2}\bar v_{4'4,58}\delta_{2'2}\delta_{3'3}\delta_{5'6}\delta_{6'7}
+\frac{1}{2}\bar v_{5'4,68}\delta_{2'2}\delta_{3'3}\delta_{4'5}\delta_{6'7}
\nonumber \\
&&
+\frac{1}{2}\bar v_{6'4,78}\delta_{2'2}\delta_{3'3}\delta_{4'5}\delta_{5'6}
-\frac{1}{2}\bar v_{2'8,24}\delta_{3'3}\delta_{4'5}\delta_{5'6}\delta_{6'7}
\nonumber \\
&&
-\frac{1}{2}\bar v_{3'8,34}\delta_{2'2}\delta_{4'5}\delta_{5'6}\delta_{6'7}
\biggr],
\label{eq-Gamma4}
\end{eqnarray}

\begin{eqnarray}
&&
\Gamma^{(3)}_{123;456}(\omega)
\nonumber \\
&=&
\frac{1}{2}\bar v_{12',3'4'}
\frac{1}{\omega-\varepsilon_{3'4'}+\varepsilon_{2'}}
\biggl[
 \frac{1}{2}\bar v_{3'2,45}\delta_{2'3}\delta_{4'6}
\nonumber \\
&+&
\frac{1}{2}\bar v_{4'3,45}\delta_{2'2}\delta_{3'6}
-\frac{1}{2}\bar v_{2'6,23}\delta_{3'4}\delta_{4'5}
\biggr].
\label{eq-Gamma3}
\end{eqnarray}

\noindent
where $\varepsilon_{123\cdots}=\varepsilon_{1}+\varepsilon_{2}+\varepsilon_{3}
+\cdots$,
and $\bar f_i=1-f_i$ with the Fermi distribution function
$f_i=f(\varepsilon_i)$.

One may wonder why in (\ref{eq-Gamma4}) and (\ref{eq-Gamma3}) 
no Fermi function factors 
appear with the propagators. This, however, is a general feature of vertices 
coupling lower configuration spaces to higher ones. A fameous example in the 
literature is given by coupling 1p1h and 2p2h spaces as this appears in the 
damping of zero sound modes, see, e.g. \cite{as89}. One may notice that the 
absence of Pauli blocking factors opens up phase space and, therefore, 
enhances the coupling.

One can ask the question whether in Eq.~(\ref{eq-fullalphaBECmass}) 
the uncorrelated mean field 3h GF should be used. One may 
think to include the hole GF's into the selfconsistent cycle or even include 
higher correlations. One should 
notice, however, that in BCS this is not done and in Eq.~(\ref{eq-bcsmass}) 
the mean field 1h GF 
is used. This has a good reason, since BCS theory is based on a variational 
wave function which fully respects the Pauli principle. Should a self 
consistent hole GF be used in (\ref{eq-bcsmass}), 
this property would be lost. We, therefore, 
also stick to the mean field hole GF's in the quartet case.

We also need the in-medium four-body Schr\"odinger equation 
for the order parameter, in analogy with the pairing case 
shown in Eq.~(\ref{eq-gapeq}) in the previous section.
It is given by 
\begin{eqnarray}
&&
\varepsilon_{1234}\langle c_{4}c_{3}c_{2}c_{1}\rangle
\nonumber \\
&&
+\sum_{1'2'3'4'} V_{1234;1'2'3'4'}\langle c_{4'}c_{3'}c_{2'}c_{1'}\rangle  =0,
\label{eq-inmedium4b}
\end{eqnarray}
where
\begin{eqnarray}
&&
V_{1234;1'2'3'4'}
\nonumber \\
&=&
(1-\rho_1-\rho_2)\frac{1}{2}\bar v_{12;1'2'} \delta_{33'}\delta_{44'}
\nonumber \\
&+&
(1-\rho_1-\rho_3)\frac{1}{2}\bar v_{13;1'3'} \delta_{22'}\delta_{44'}
\nonumber \\
&+&
{\rm permutations}.
\end{eqnarray}
Details  are given in \cite{rss98} 
and Appendix~\ref{sec-appendix-inmedium}.

We consider symmetric (nuclear) matter.
In this case, 
we can give a fully symmetric order parameter of exchange between 
two particles with respect to momenta:
\begin{eqnarray}
\langle c_{4}c_{3}c_{2}c_{1}\rangle \to 
\phi_{\vec k_1,\vec k_2,\vec k_3,\vec k_4} 
\chi_0,
\label{eq-orderparameterphi}
\end{eqnarray} 
where the  spin-isospin singlet wave function is represented by $\chi_0$,
and we consider here a spin-isospin independent two-body interaction:
\begin{eqnarray}
\bar v_{12,34}
&\to&
v_{\vec k_1 \vec k_2, \vec k_3 \vec k_4}
(\delta_{s_1s_3}\delta_{t_1t_3}
 \delta_{s_2s_4}\delta_{t_2t_4}
\nonumber \\
&&\qquad\qquad
-\delta_{s_1s_4}\delta_{t_1t_4}
 \delta_{s_2s_3}\delta_{t_2t_3})
\label{eq-twobodyinteraction}
\end{eqnarray}
with $s_i$ ($t_i$) of spin (isospin) index. 
$v_{\vec k_1 \vec k_2, \vec k_3 \vec k_4}$ 
is symmetric with respect to
exchange of the momenta: 
$v_{\vec k_1 \vec k_2, \vec k_3 \vec k_4}
=v_{\vec k_2 \vec k_1, \vec k_3 \vec k_4}
=v_{\vec k_1 \vec k_2, \vec k_4 \vec k_3}$.

\begin{figure}
\includegraphics[width=75mm]{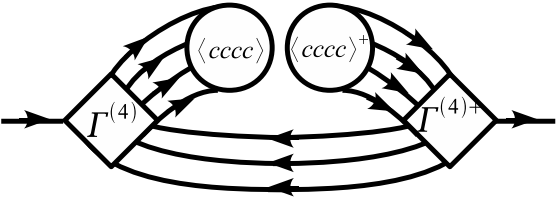}
\caption{\label{fig-alphaBECmass}Graphic representation of the 
mass operator for alpha condensation
in Eq.~(\ref{eq-fullalphaBECmass}).}
\end{figure}

Then, Eq.~(\ref{eq-inmedium4b}) is explicitly written as
\begin{eqnarray}
&&
\sum_{i=1}^4\varepsilon_i
\phi_{\vec k_1 \vec k_2 \vec k_3 \vec k_4}
+\int \prod_{i=1}^4\frac{d^3k_i'}{(2\pi)^3}
\nonumber \\
&\times&
\biggl[
 (1-\rho(\vec k_1)-\rho(\vec k_2))v_{\vec k_1\vec k_2,\vec k_1'\vec k_2'}
\nonumber \\
&&\qquad
\times
 (2\pi)^3\delta(\vec k_3-\vec k_3')(2\pi)^3\delta(\vec k_4-\vec k_4')
\nonumber \\
&&
+(1-\rho(\vec k_1)-\rho(\vec k_3))v_{\vec k_1\vec k_3,\vec k_1'\vec k_3'}
\nonumber \\
&&\qquad
\times
 (2\pi)^3\delta(\vec k_2-\vec k_2')(2\pi)^3\delta(\vec k_4-\vec k_4')
\nonumber \\
&&
+\mbox{permutations}
\biggr]\phi_{\vec k_1' \vec k_2' \vec k_3' \vec k_4'}
=0.
\label{eq-inmedium4BSE}
\end{eqnarray}
\noindent
A sketch of the quartet mass operator is shown 
in Fig.~\ref{fig-alphaBECmass}.

\section{\label{sec-approximate}approximate 
quartet mass operator}

\begin{figure}
\includegraphics[width=75mm]{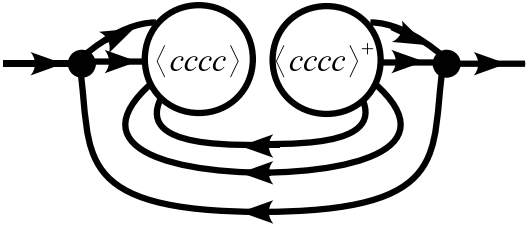}
\caption{\label{fig-alphamassapp}The graphical representation of 
the approximate $\alpha$-BEC mass operator $M^{\rm quartet}$ 
of Eq.~(\ref{eq-Qmassoperator}). }
\end{figure}

From Eqs.~(\ref{eq-Gamma4}) and (\ref{eq-Gamma3}) it 
becomes evident that the full evaluation of
the 1p $\leftrightarrow$ 4p3h vertices
are too complicated to be evaluated exactly.
However, with quite reasonable approximations, 
one arrives at a numerically manageable expression.
Since this discussion involves lengthy, quite technical details, 
we relegate it 
to Appendix~\ref{sec-appendix-discussion}
and only give the final result here:
\begin{eqnarray}
M^{\rm quartet}_{1;1}(\omega)=\sum_{234} 
\frac{\tilde \Delta_{1234}(\bar f_2 \bar f_3 \bar f_4+ f_2 f_3 f_4)
\tilde \Delta_{1234}^*}{\omega+\varepsilon_{234}}
\label{eq-Qmassoperator}
\end{eqnarray}
where the quartet `gap' matrix $\tilde \Delta_{1234}$ is given by
\begin{eqnarray}
\tilde \Delta_{1234}
=
\lambda' \frac{1}{2}\bar v_{12,1'2'}\delta_{33'}\delta_{44'}
\langle c_{1'} c_{2'} c_{3'} c_{4'} \rangle.
\label{eq-tildedelta}
\end{eqnarray}
The graphical representation of the approximate $M^{\rm quartet}$ is shown 
in Fig.~\ref{fig-alphamassapp}.

In Eq.~(\ref{eq-tildedelta}), 
we put a renormalization factor $\lambda'$ in front of the 
vertex which shall effectively account for the approximations  considered, 
see App.~\ref{sec-appendix-discussion}. 
It is, however, very fortunate that the 
final result is 
independent of the value of $\lambda'$ and, therefore, one also 
may drop it. This somewhat surprising effect is due to selfconsistency and 
self-readjustment of the solution. 
It is demonstrated in the Appendix~\ref{sec-appendix-parameter} for 
the simple case of ordinary pairing.

The complexity of the 
calculation still is further much reduced 
for the order parameter (\ref{eq-orderparameterphi})
with our mean field ansatz projected on zero total momentum, as already very
successfully employed in~\cite{slr09}
\begin{eqnarray}
\phi_{\vec k_1 \vec k_2,\vec k_3 \vec k_4}
&=& 
\varphi(|\vec k_1|)\varphi(|\vec k_2|)\varphi(|\vec k_3|)\varphi(|\vec k_4|)
\nonumber \\
&&\times 
(2\pi)^3\delta(\vec k_1+\vec k_2+\vec k_3+\vec k_4).
\label{eq-PHF4bwf}
\end{eqnarray}
It should be pointed out that this product ansatz 
with four identical $0S$ single particle wave functions is typical for a 
ground state configuration of the quartet. Excited configurations with wave 
functions of higher nodal structures may eventually be envisaged for other 
physical situations. We also 
would like to mention that the momentum conserving $\delta$ function induces 
strong correlations among the four particles and is, therefore, a rather non 
trivial variational wave function.

For the two-body interaction of $v_{\vec k_1\vec k_2,\vec k_3 \vec k_4}$ 
in Eq.~(\ref{eq-twobodyinteraction}), we employ the 
same separable form as done already in our previous publication on the 
quartet critical temperature in Ref.~\cite{slr09}:
\begin{eqnarray}
v_{\vec k_1\vec k_2,\vec k_3\vec k_4} 
&=& 
\lambda 
w(\frac{\vec k_1-\vec k_2}{2})
w(\frac{\vec k_3-\vec k_4}{2})
\nonumber \\
&&\times 
(2\pi)\delta^3(\vec k_1+\vec k_2-\vec k_3-\vec k_4)
\end{eqnarray}
with the form factor $w(\vec k) = w(|\vec k|) = e^{-k^2/b^2}$. 
An example for strength and range parameters are given
in Sec.~\ref{sec-result} below.

With these simplifications,
the mass operator (\ref{eq-Qmassoperator}) is independent of 
spin and isospin, and therefore 
it can be reduced to the following four-fold integral
\begin{widetext}
\begin{eqnarray}
&&
M^{\rm quartet}(k_1,\omega)
\nonumber \\
&=&
\frac{1}{(4\pi^2)^4}
\int^{\infty}_{0} \!\!\! dKK^2 \int^{1}_{-1} \!\!\! dt_1
\int^{\infty}_{0} \!\!\! dk k^2 \int^{1}_{-1}\!\!\! dt
\frac{\bar f(|\vec K-\vec k_1|)
      \bar f(|\frac{\vec K}{2}+\vec k|)
      \bar f(|\frac{\vec K}{2}-\vec k|)
     +f(|\vec K-\vec k_1|) f(|\frac{\vec K}{2}+\vec k|) 
      f(|\frac{\vec K}{2}-\vec k|)}
     {\omega+\varepsilon_{\vec K-\vec k_1}
            +\varepsilon_{\frac{\vec K}{2}+\vec k}
            +\varepsilon_{\frac{\vec K}{2}+\vec k}}
\nonumber \\
&\times&
(w(|\vec k_1-\frac{\vec K}{2}|))^2 
(\varphi(|\frac{\vec K}{2}+\vec k|))^2(\varphi(|\frac{\vec K}{2}-\vec k|))^2
\left[
\int^{\infty}_{0}\!\!\! dk'k'^2 \int^{1}_{-1}\!\!\! dt'
w(k')\varphi(|\frac{\vec K}{2}+\vec k'|)
\varphi(|\frac{\vec K}{2}-\vec k'|)\right]^2,
\label{eq-appqmo}
\end{eqnarray}
\end{widetext}
with $t_1=(\vec K\cdot \vec k_1)/(Kk_1)$,
$t=(\vec K\cdot \vec k)/(Kk)$, and 
$t'=(\vec K\cdot \vec k')/(Kk')$.
Here we represented the Fermi distribution function as
$f_1\to f(k_1)=f(\varepsilon_1)$.
In this expression any strength factor of the vertices has been dropped, 
see our above argument and Appendix~\ref{sec-appendix-parameter}. 
For the imaginary part of this expression an energy conserving 
delta function comes instead of the full denominator and then 
the 4D integral can be reduced to a 3D one. 
How this goes in detail is again explained 
in Appendix~\ref{sec-appendix-Immass}. 
The real part of $M^{\rm quartet}$ is then obtained from the 
imaginary part via a dispersion integral:
\begin{eqnarray}
&&
{\rm Re}M^{\rm quartet}(k,\omega+i\eta)
\nonumber \\
&=&
-{\cal P}\int^{\infty}_{-\infty}\!\!\!
\frac{d\omega'}{\pi}
\frac{{\rm Im}M^{\rm quartet}(k,\omega'+i\eta)}{\omega-\omega'}
\label{eq-remassoperator}
\end{eqnarray}
where ${\cal P}$ denotes the Cauchy principal value.

The occupation numbers are finally obtained from
\begin{widetext}
\begin{eqnarray}
\rho(k)
&=&
-\int \frac{d\omega}{2\pi}
2{\rm Im}G(k,\omega+i\eta)f(\omega)
\nonumber \\
&=&
\int \frac{d\omega}{2\pi}
\frac{-2{\rm Im}M^{\rm quartet}(k,\omega+i\eta)}
{(\omega-\varepsilon({\vec k})-{\rm Re}M^{\rm quartet}(k,\omega+i\eta))^2
 +({\rm Im}M^{\rm quartet}(k,\omega+i\eta))^2}f(\omega).
\label{eq-rho}
\end{eqnarray}
\end{widetext}

The equation for the order parameter (10) is formally not changed from 
Eqs.~(4)-(7) of \cite{slr09} 
but the occupation numbers are calculated selfconsistently with 
above equation. For completeness we again give the equations for the single 
particle wave function $\varphi(k)$

\begin{widetext}
\begin{eqnarray}
{\cal A}(k)\varphi(k)+3{\cal B}(k)+3{\cal C}(k)\varphi(k)=0,
\label{eq-PHFequation}
\end{eqnarray}

where

\begin{eqnarray}
{\cal A}(k)
&=&
\int \frac{d^3 k_{2}}{(2\pi)^3}\frac{d^3 k_{3}}{(2\pi)^3}
     \frac{d^3 k_{4}}{(2\pi)^3}
\Bigl(
\frac{k^2}{2m}+\frac{k^2_{2}}{2m}+\frac{k^2_{3}}{2m}+\frac{k^2_{4}}{2m}-4\mu
\Bigr)
\varphi^2(k_{2})\varphi^2(k_{3})\varphi^2(k_{4})
(2\pi)^3\delta(\vec k+\vec k_{2}+\vec k_{3}+\vec k_{4})
\nonumber \\
&=&
\frac{1}{(2\pi)^4}\int dKK^2 
\int^{1}_{-1}dt
\Bigl(
\frac{k^2}{2m}+\frac{3}{2m}P^2
-4\mu
\Bigr)
\varphi^2(P)
\int dk' k'^2\int^{1}_{-1} dt'
\varphi^2(k')
\varphi^2(P'),
\label{eq-fourbody-a}
\end{eqnarray}

\begin{eqnarray}
{\cal B}(k)
&=&
\int \frac{d^3 k_{2}}{(2\pi)^3}\frac{d^3 k_{3}}{(2\pi)^3}
\frac{d^3 k_{4}}{(2\pi)^3}
\frac{d^3k_{1}'}{(2\pi)^3}\frac{d^3k_{2}'}{(2\pi)^3}
(1-\rho(k)-\rho(k_{2}))
\lambda w(\frac{\vec k-\vec k_{2}}{2})w(\frac{\vec k_{1}'-\vec k_{2}'}{2}) 
(2\pi)^3\delta(\vec k+\vec k_{2}-\vec k_{1}'-\vec k_{2}')
\nonumber \\
&&\times 
\varphi(k_{1}')\varphi(k_{2})\varphi(k_{2}')\varphi^2(k_{3})\varphi^2(k_{4})
(2\pi)^3\delta(\vec k+\vec k_{2}+\vec k_{3}+\vec k_{4})
\nonumber \\
&=&
\frac{\lambda}{(2\pi)^6}\int dKK^2 \int^{1}_{-1} dt 
\Bigl(1-\rho(k)-\rho(P)\Bigr)
w(Q)
\varphi(P)
\int dk'k'^2 \int^{1}_{-1} dt'
w(Q')
\varphi(k')\varphi(P')
\nonumber \\
&&\times
\int dk''k''^2 \int^{1}_{-1} dt''
\varphi^2(k'')
\varphi^2(P''),
\label{eq-fourbody-b}
\end{eqnarray}

\begin{eqnarray}
{\cal C}(k)
&=&
\int \frac{d^3 k_{2}}{(2\pi)^3}\frac{d^3 k_{3}}{(2\pi)^3}
\frac{d^3 k_{4}}{(2\pi)^3}
 \frac{d^3k_{2}'}{(2\pi)^3}\frac{d^3k_{3}'}{(2\pi)^3}
(1-\rho(k_{2})-\rho(k_{3})) \lambda
w(\frac{\vec k_{2}-\vec k_{3}}{2})w(\frac{\vec k_{2}'-\vec k_{3}'}{2})
(2\pi)^3\delta(\vec k_{2}+\vec k_{3}-\vec k_{2}'-\vec k_{3}')
\nonumber \\
&&\times 
\varphi(k_{2})\varphi(k_{2}')\varphi(k_{3})\varphi(k_{3}')\varphi^2(k_{4})
(2\pi)^3\delta(\vec k+\vec k_{2}+\vec k_{3}+\vec k_{4})
\nonumber \\
&=&
\frac{\lambda}{(2\pi)^6}
\int dKK^2 
\int^{1}_{-1} dt \varphi^2(P) 
\int dk'k'^2 \int^{1}_{-1} dt'
\Bigl(1-\rho(k')-\rho(P')\Bigr)
w(Q') 
\varphi(k')\varphi(P')
\nonumber \\
&&\times
\int dk''k''^2 \int^{1}_{-1} dt''
w(Q'')
\varphi(k'')\varphi(P'').
\label{eq-fourbody-c}
\end{eqnarray}
\end{widetext}
with 
\begin{eqnarray*}
P&=&\sqrt{K^2+k^2+2Kkt},
\\
P'&=&\sqrt{K^2+k'^2+2Kk't'},
\\
P''&=&\sqrt{K^2+k''^2+2Kk''t''},
\\
Q&=&\sqrt{K^2/4+k^2+Kkt},
\\
Q'&=&\sqrt{K^2/4+k'^2+Kk't'},
\\
Q''&=&\sqrt{K^2/4+k''^2+Kk''t''}.
\end{eqnarray*}

As mentioned, in these equations the occupation numbers $\rho(k)$ shall be 
calculated selfconsistently from Eq.~(\ref{eq-rho}).

Because of its particular importance, before the presentation of the results, 
we first will discuss in the following section the 3h level density.

\section{\label{sec-leveldensity}Three hole level density}

\begin{figure}
\includegraphics[width=55mm]{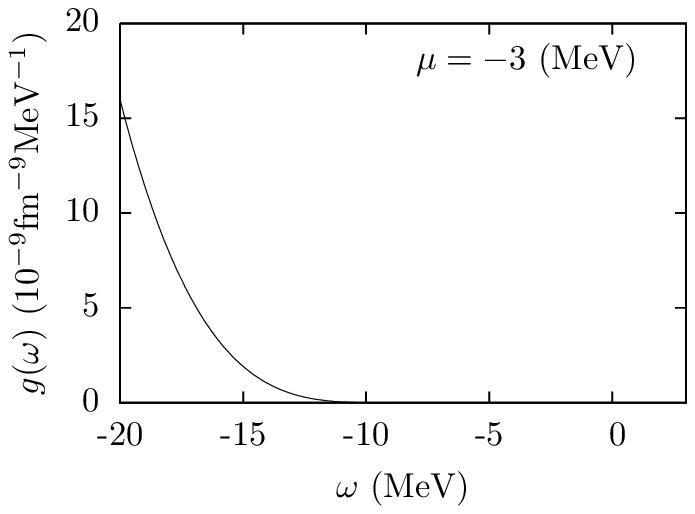}
\includegraphics[width=55mm]{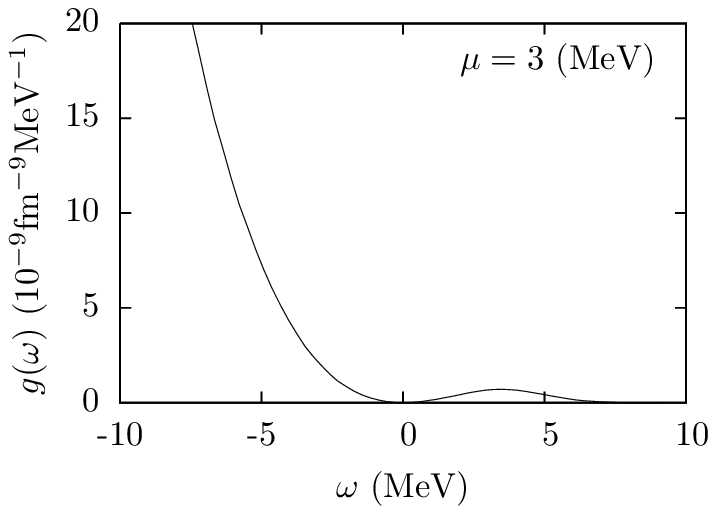}
\includegraphics[width=55mm]{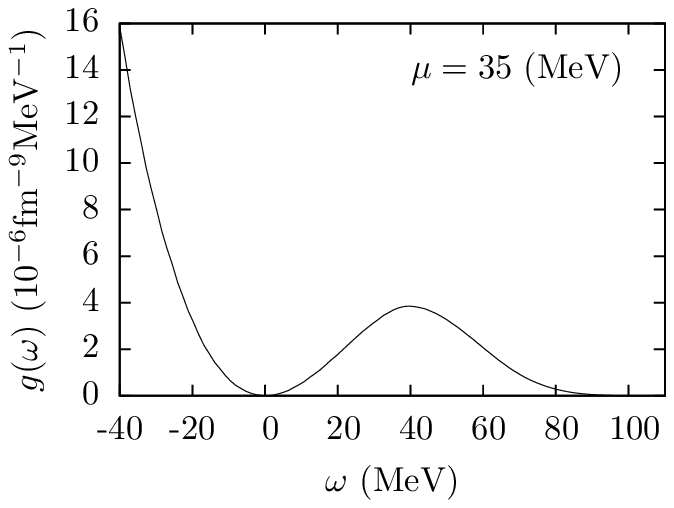}
\caption{\label{fig-ld}
3h evel densities defined in Eq.~(\ref{eq-ld}) for various values of the 
chemical potential $\mu$ at a zero temperature.}
\end{figure}

In what follows a crucial role will be played by the 3 hole propagator 
entering the mass operator. Since its influence on quartet condensation 
will be radically different from the corresponding 1h propagator 
in the pairing case, we will pay special attention to it in this section. 
In mean field approximation we can write

\begin{eqnarray}
\frac{(\bar f_{1}\bar f_{2}\bar f_{3}+f_{1}f_{2}f_{3})}
     {\omega+\varepsilon_{123}}
=
G^{(3h)}(k_1,k_2,k_3;\omega),
\end{eqnarray}

\noindent
where at r.h.s. we dropped spin and isospin indices 
since we consider unpolarised (nuclear) matter.
We immediately see a strong difference with the pairing case.
There, only a single hole line enters whose numerator is 
(see previous section) 
$\bar f_1+f_1=1$, see Eq.~(\ref{eq-bcsmass})
Therefore, no single particle occupation numbers appear in the numerator
of a single hole propagator. This difference between the three hole and one
hole propagators leads to strong consequences. This is best demonstrated 
in Fig.~\ref{fig-ld} with the three hole level density

\begin{eqnarray}
&&
g(\omega)
\nonumber \\
&=&-\int \frac{d^3k_1}{(2\pi)^3}\frac{d^3k_2}{(2\pi)^3}\frac{d^3k_3}{(2\pi)^3}
{\rm Im} G^{(3h)}(k_1,k_2,k_3;\omega+i\eta)
\nonumber \\
&=&\int \frac{d^3k_1}{(2\pi)^3}\frac{d^3k_2}{(2\pi)^3}\frac{d^3k_3}{(2\pi)^3}
\nonumber \\
&&\times
(\bar f_1 \bar f_2 \bar f_3+ f_1 f_2 f_3)
\pi\delta(\omega+\varepsilon_1
                +\varepsilon_2
                +\varepsilon_3)
\nonumber \\
&=&
\frac{m}{(2\pi)^5}
\int^{k_{\rm max}}_{0} \!\!\!\!\!\!\!\!\!\!\!\! dk k^2 
\int^{K_{\rm max}}_{0} \!\!\!\!\!\!\!\!\!\!\!\! dK K^2 p
\nonumber \\
&&\times
\Bigl(
\bar f(k)\bar F(K,p)
+ f(k)F(K,p)\Bigr),
\label{eq-ld}
\end{eqnarray} 
where
\begin{eqnarray}
k_{\rm max}
&=&
\sqrt{2m(3\mu-\omega)},
\label{eq-kmax}
\\
K_{\rm max}
&=&
\sqrt{4m(3\mu-\omega)-2k^2},
\\
p
&=&
\sqrt{3m\mu-m\omega-\frac{k^2}{2}-\frac{K^2}{4}},
\end{eqnarray}
and 
\begin{eqnarray}
F(K,p)
&=&
\int^1_{-1} dt 
f(|\frac{\vec K}{2}+\vec p|)f(|\frac{\vec K}{2}-\vec p|),
\\
\bar F(K,p)
&=&
\int^1_{-1} dt 
\bar f(|\frac{\vec K}{2}+\vec p|)\bar f(|\frac{\vec K}{2}-\vec p|)
\end{eqnarray}
with $t=\vec K\cdot\vec p/(Kp)$.

In Fig.~\ref{fig-ld} 
we show the level density at zero temperature 
($f(\omega)=\theta(-\omega)$), where it is calculated with
the proton mass $m=938.27$MeV (natural unit).
Two cases have to be considered, chemical 
potential $\mu$ positive or negative. In the latter case we have binding 
of the quartet. Let us first discuss the case $\mu>0$. 
We remark that in this case, the 3h level density goes through
zero at $\omega=0$, i.e. just in the region where the quartet correlations 
should appear. This is a strong difference with the pairing case where the 
1h level density does not feel any influence from the medium and, therefore, 
the corresponding level density varies (neglecting the mean field for the sake 
of the argument) like in free space with the square root  of energy. 
In particular, 
this means that the level density is {\it finite} at the Fermi level. 
This is a 
dramatic difference with the quartet case and explains why Cooper pairs can 
strongly overlap whereas for quartets this is impossible as we will see below.
We also would like to point out that the 3h level density is just the 
mirror to the 3p level density which has been discussed in ~\cite{bhh86}. 

For the case where $\mu<0$ there is nothing very special, 
besides the fact that it only is non-vanishing 
for negative values of $\omega$ and that the upper 
boundary is given by $\omega = 3\mu$.
Therefore, the level density of Eq.~(\ref{eq-ld})
is zero for $\omega>3\mu$.

\section{\label{sec-result}Results and discussion}

\begin{figure}
\includegraphics[width=27mm]{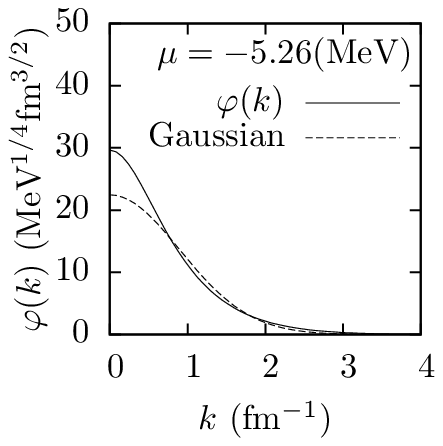}
\includegraphics[width=28.5mm]{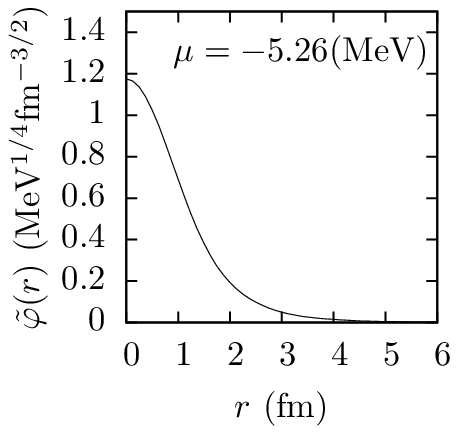}
\includegraphics[width=29mm]{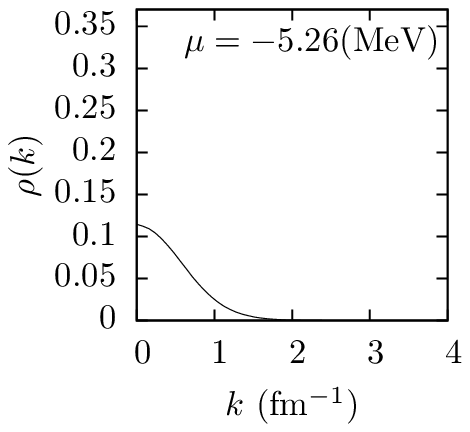}

\includegraphics[width=27mm]{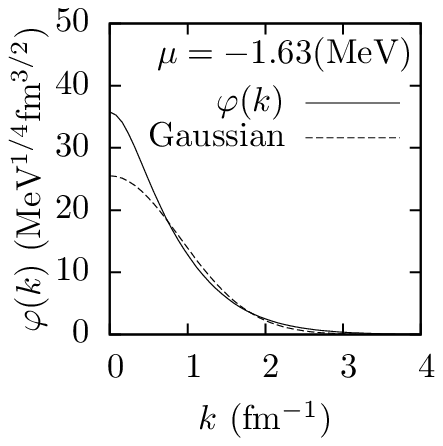}
\includegraphics[width=28.5mm]{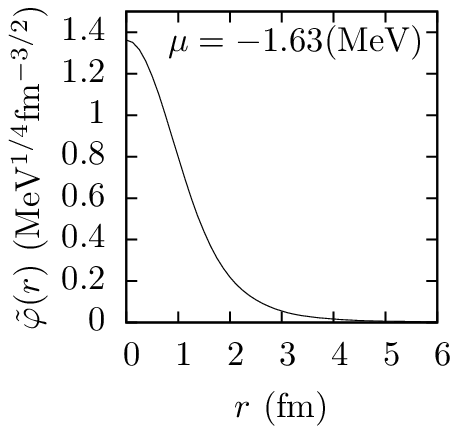}
\includegraphics[width=29mm]{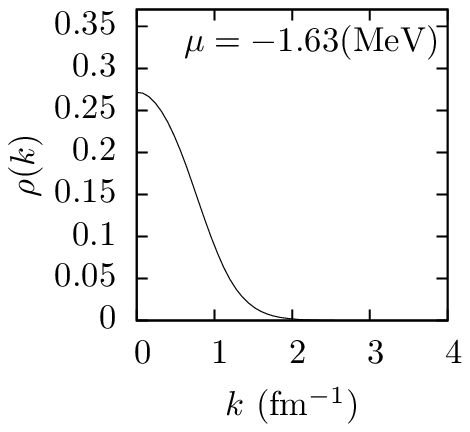}

\includegraphics[width=27mm]{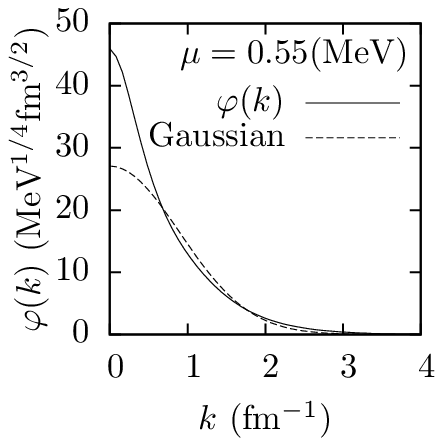}
\includegraphics[width=28.5mm]{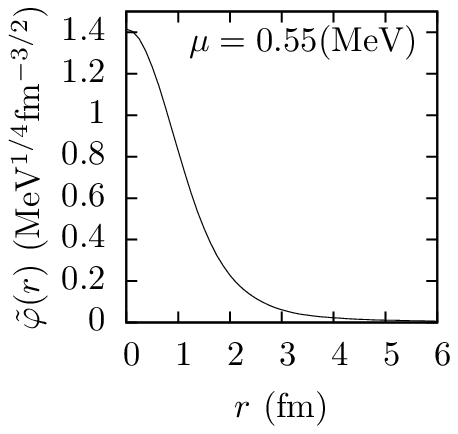}
\includegraphics[width=29mm]{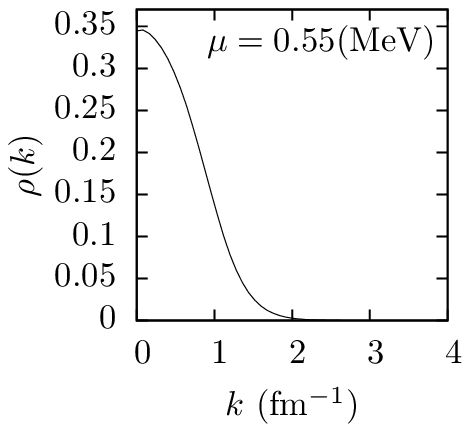}
\caption{\label{fig-spwf}
Single particle wave function $\varphi(k)$ in $k$-space (left),
for $r$-space $\tilde \varphi(r)$ (middle), and  
occupation numbers (right)
at $\mu=-5.26$ (top), $-1.63$ (middle) and $0.55$ (bottom).
The $r$-space wave function $\tilde \varphi(r)$
is derived from the Fourier transform of $\varphi(k)$
by $\tilde \varphi(r)=\int d^3k e^{i\vec k \cdot \vec r}\varphi(k)/(2\pi)^3$.
The dashed line in the left panels correspond to the Gaussian 
with same norm and rms momentum as $\varphi(k)$.}
\end{figure}

\begin{figure}
\includegraphics[width=60mm]{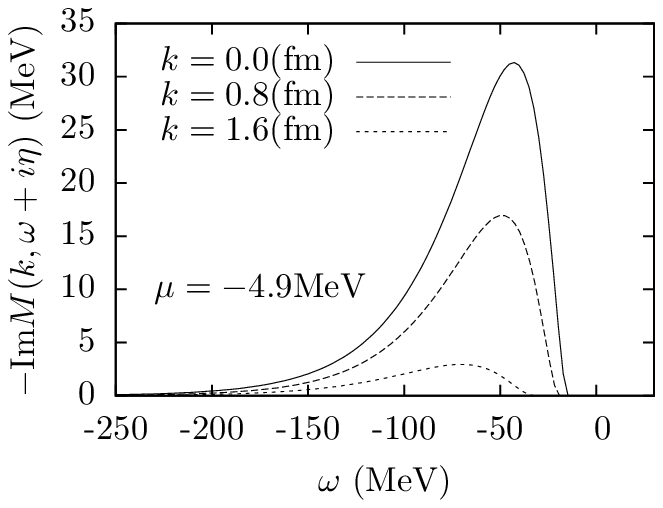}
\includegraphics[width=60mm]{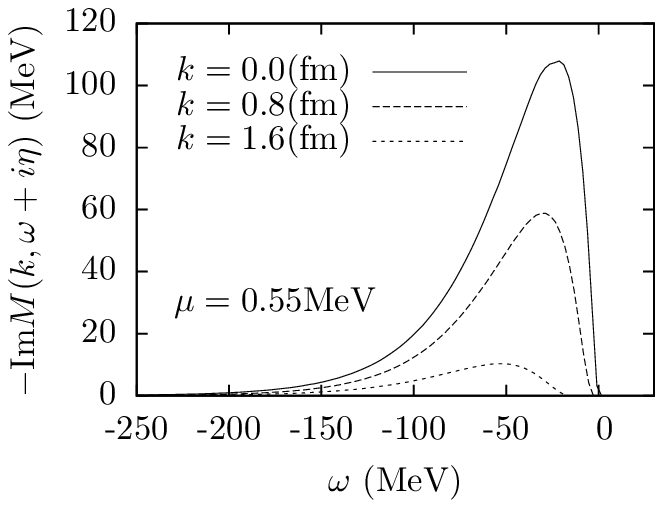}
\caption{\label{fig-immassoperator}
$-$Im$M^{\rm quartet}(k_1,\omega+i\eta)$ in Eq.~(\ref{eq-appqmo})
as a function of $\omega$ for $\mu=-4.9$MeV (left) 
and for $\mu=0.55$MeV (right)  at  zero temperature.}
\end{figure}

\begin{figure}
\includegraphics[width=60mm]{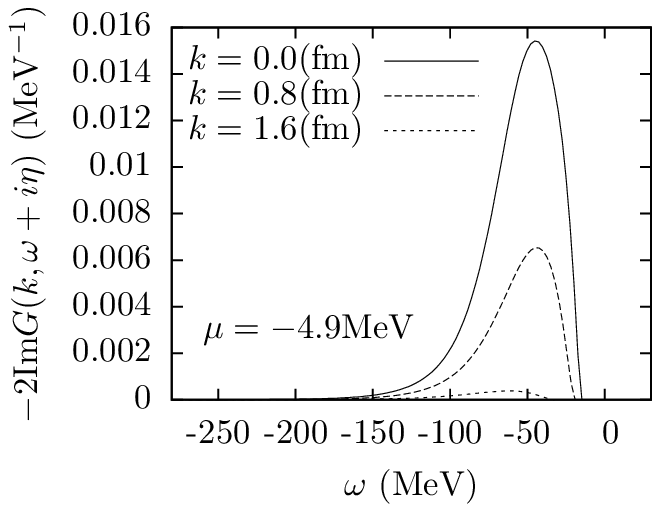}
\includegraphics[width=60mm]{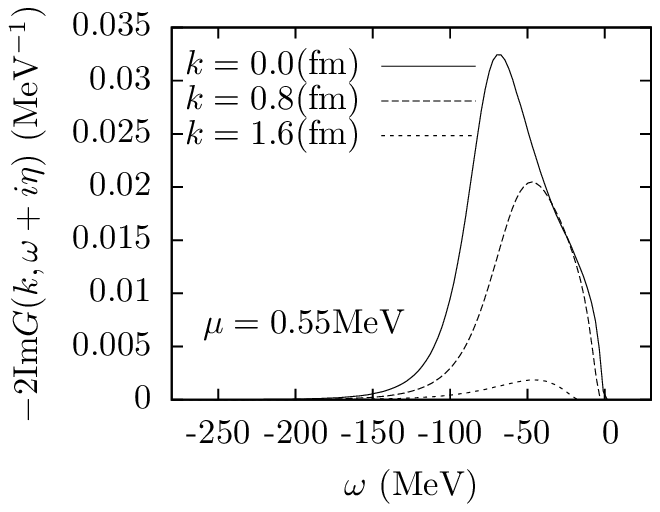}
\caption{
\label{fig-2ImG11}
$-2$Im$G(k,\omega+i\eta)$ in Eq.~(\ref{eq-rho}) 
as function of $\omega$ 
for $\mu=-4.9$MeV (top) 
and for $\mu=0.55$MeV (bottom) at a zero temperature.}
\end{figure}

At first let us mention that in this pilot application of 
our selfconsistent quartet theory,
we only will consider the zero temperature case. As a definite physical 
example, we will treat the case of nuclear physics with the particularly 
strongly bound quartet, the $\alpha$ particle. It should be pointed out, 
however, that if scaled appropriately all energies and lengths can be 
transformed to other physical systems. For the nuclear case it is 
convenient to measure energies in in Fermi energies $\varepsilon_F = 35$MeV 
and lengths in inverse Fermi momentum $k_F^{-1} = 1.35^{-1}$fm.

To determine the order parameter, 
the calculation iterates in the following cycle:

\begin{enumerate}
\item     Fix a chemical potential $\mu$
\item     Give an initial trial order parameter $\varphi(k)$
\item     Calculate the mass operator $M^{\rm quartet}(k,\omega)$ 
          of Eq.~(\ref{eq-appqmo}) with $\varphi(k)$
\item     Calculate the occupation numbers $\rho(k)$ 
          with $M^{\rm quartet}(k,\omega)$ from Eq.~(\ref{eq-rho})
\item     Substituting $\varphi(k)$ and $\rho(k)$ into 
          Eqs.~(\ref{eq-fourbody-a}), (\ref{eq-fourbody-b}), and 
          (\ref{eq-fourbody-c}),
          derive the ``new'' single particle wave function $\varphi(k)$ from
\begin{eqnarray}
\varphi(k)=\frac{-3{\cal B}(k)}{{\cal A}(k)+3{\cal C}(k)}.
\label{eq-varphi}
\end{eqnarray}
\item     Quit the cycle once $\varphi(k)$ has converged, while go to 3. 
otherwise.
\end{enumerate}

The single particle wave functions and occupation numbers 
obtained from the above cycle 
are shown in Fig.~\ref{fig-spwf}. We take $\lambda=-992$MeV fm$^3$
and $b=1.43$fm$^{-1}$ to get the binding energy of 
the free $\alpha$-particle ($-28.3$MeV) and its radius 
($1.7$fm).
We also insert the Gaussian wave function with same rms momentum
as the single particle wave function in the left figures 
in Fig.~\ref{fig-spwf}.
As shown in Fig.~\ref{fig-spwf},
the single particle wave function is sharper than a
Gaussian. There is the term ${\cal A}(k)$ of Eq.~(\ref{eq-fourbody-a})
in the denominator of $\varphi(k)$ in Eq.~(\ref{eq-varphi}),
and ${\cal A}(k)$  involves the factor $(k^2/(2m)-4\mu)$.
Hence, $\varphi(k)$ is closer to a Lorentzian rather 
than to a Gaussian~\cite{alphaBECfinite}.

We could not obtain a convergent wave function
for $\mu>0.55$MeV. This difficulty is of the same origin as in the case of our 
calculation of the critical temperature for $\alpha$ particle condensation.
This stems from the fact that 
for  larger positive values of the chemical potential,
the denominator of $\varphi(k)$ in Eq.~(\ref{eq-varphi})
at a certain value of $k$ becomes zero, 
while the numerator is finite.
In case of pair condensation, as shown in Eq.~(\ref{eq-gapeq}),
both the denominator and numerator become zero at the same value of  $k$, 
this being a further crucial difference between the quartet and pairing cases.
In the r.h.s. panels of Fig.~\ref{fig-spwf}
we also show the corresponding occupation numbers. 
We see that they are very small. However, they increase for increasing values 
of the chemical potential. For $\mu = 0.55$ MeV the maximum of the occupation 
still only attains 0.35 what is far away from the saturation value of one. 
What really happens for larger values of the chemical potential, is unclear. 
Surely, as discussed in  Sec.~\ref{sec-leveldensity} the situation for the 
quartet case is completely different from the standard pairing case. This 
is due to the fact, as already mentioned, 
that the 3h level density goes through zero at 
$\omega=0$, i.e. just at the place where the quartet correlation should 
build up for positive values of $\mu$. Due to this fact, 
the inhibition to go into the positive $\mu$ regime is here even stronger 
than in the case of the critical temperature~\cite{slr09}.

The situation in the quartet case is also in so far much different, 
as the the 3h GF produces a considerable imaginary part of the mass operator.
Figure~\ref{fig-immassoperator} shows the imaginary part of 
the approximate quartet mass operator
of Eq.~(\ref{eq-appqmo}) 
for $\mu<0$ and $\mu>0$.
These large values of the damping rate imply a strong violation of the 
quasiparticle picture. In Fig.~\ref{fig-2ImG11} 
we show the spectral function of the single 
particle GF. Contrary to the pairing case with its sharp quasiparticle pole, 
we here only find a very broad distribution, implying that the quasiparticle 
picture is completely destroyed. How to formulate a theory which goes 
continuously from the quartet case into the pairing case, is an open question. 
One solution could be to start right from the beginning with an in medium four 
body equation which contains a superfluid phase. When the quartet phase 
disappears, the superfluid phase may remain. Such investigations shall be 
done in the future.

\section{conclusion}

We formulated the gap equation for quartetting in fermion systems
in analogy to the BCS gap equation.
The mass operator of quartet-BEC with effective 
four-body vertices and in-medium four-body Schr\"odinger equation
was derived  with the Dyson equation approach to correlation functions.
However, the full expression of the quartet mass operator 
is too complicated to be evaluated numerically in a direct manner. 
The biggest problem stems from 
the many dimensional integrals over momenta.
We, therefore, introduced some reasonable approximations reducing the 
complexity considerably. A feature of great help is that the final 
answer is independent of the strength of the vertex function, making our 
approximation probably quite reliable. In our calculation we also applied 
the same mean field ansatz projected on zero total momentum which was already 
so successful in our previous calculation 
of the critical temperature~\cite{slr09}. 
This feature, of course, reduces the numerical effort tremendously, since 
only a single 0S wave function has to be determined selfconsistently by 
iteration.

In this pilot work with an application to nuclear physics, we showed  
results only at zero temperature, however, the formalism we presented is 
at finite temperature.

We think that the results are of general validity showing qualitatively very 
distinct features from the pairing case. For example no well defined 
quasiparticles occur in the case of quartets. This is due to the fact that 
the quartet order parameter in the single particle mass operator goes along 
with the level density of three uncorrelated holes. Only the total momentum of 
the three holes is well defined and equal to the time reversed momentum of the 
incoming particle. The relative momenta of the three holes have to be 
integrated over, yielding a strong imaginary part of the mass 
operator smoothing out any individual single particle structure.
Another remarkable feature already encountered in our previous work 
in Ref.~\cite{slr09} 
is that the self consistent solution 
seems to exist only from negative chemical potential until around zero,
i.e. from the BEC, or strong coupling  region
until crossover region.
Once one goes to positive $\mu$'s the 
solution breaks down. This effect is even more pronounced here than it was in 
our study \cite{slr09} for the critical temperature. This can be traced 
back to the fact that the 
3h level density goes through zero at 3$\mu$ for $\mu>0$, that is just at the 
place where the quartet correlations should occur. Actually this feature is 
present for all multi-particle multi-hole level densities. The only exception 
being the single particle level density which is finite at the Fermi level. 
This unique feature makes that pairing also is unique and for instance allows 
for a weak coupling situation with a coherence length of the pair orders of 
magnitude larger than the interparticle distance. It remains an open problem, 
how to formulate a more general theory 
which continuously goes from the quartet case to 
the pairing case.

\acknowledgments
P. S. wants to thank M. Urban for useful discussions.
This work is supported by the DFG Grant No. RO905/29-1.

\appendix

\section{ Dyson equations for multi-particle-hole Green functions}

We shall shortly review  the basic formulation used in the present work. 
We here extend the approach 
for  real-time Green's functions (GF's) at  zero temperature 
in Ref.~\cite{drs98}
to real-time GF's at  finite temperature~\cite{fw,mahan}.

The Hamiltonian in a fermion system with two-body interaction is

\begin{eqnarray}
&&
K
=H-\mu N=T+V-\mu N
\nonumber \\
&=&
\sum_{1}\varepsilon_1 c_1^\dagger c_1
+\frac{1}{4}\sum_{1234}\bar v_{12,34} c_1^\dagger c_2^\dagger c_4 c_3.
\end{eqnarray}
The $c_1$, $c_1^\dagger$ are fermion annihilation and creation operators
with an arbitrary quantum number $1$.

A real-time multi-particle-hole GF 
at a finite temperature is defined by
\begin{eqnarray}
&&
G^{(i{\rm p}j{\rm h})}_{\alpha;\alpha'}(t-t')
\nonumber \\
&=&
\left\{
\begin{array}{ll}
-i\langle  T(A_{\alpha}(t) A_{\alpha'}^\dagger(t'))  \rangle
& \mbox{(chronological)} \\
-i\theta(t-t')\langle  [A_{\alpha}(t),A_{\alpha'}^\dagger(t')]_{\pm}  \rangle
& \mbox{(retarded)} 
\end{array}
\right.
\label{eq-app-ipjhGF}
\end{eqnarray}
where $\langle ...\rangle$ means the thermal average, 
$T$ is the time ordering operator, 
$A_{\alpha}$ is an arbitrary operator consisting out of individual fermion 
operators $c_{1}$ and $c_{1}^\dagger$,
and $[...,...]_{\pm}$ is the anti-commutator or commutator.
The time dependence of the operators is given in the Heisenberg picture
$A_{\alpha}(t)=e^{iKt}A_{\alpha}e^{-iKt}$.

Note that, although we treat chronological GF's below, 
the change to retarded, advanced and Matsubara GF's 
goes as usual~\cite{fw}.
Below, when we go from time space to Fourier space,
we always will go over to retarded GF's 
without mentioning it explicitly.

The superscript ($i$p$j$h) in Eq.~(\ref{eq-app-ipjhGF})
means $i$-particle $j$-hole GF, where
$i$ ($j$) is the number of the annihilation (creation) operators 
in $A_{\alpha}$, e.g.  
$G^{(1{\rm p})}_{1;1'}(t-t')
=-i\langle T(c_{1}(t) c_{1'}^\dagger(t'))  \rangle$,
$G^{(1{\rm p}1{\rm h})}_{1,2;1',2'}(t-t')
=-i\langle  T((c_{2}^\dagger c_{1})_t 
(c_{1'}^\dagger c_{2'})_{t'})  \rangle$
with $(c_{2}^\dagger c_{1})_t=
c_{2}^\dagger(t) c_{1}(t)$, etc..

The Dyson equation for the $i$p$j$h GF is \cite{drs98} 
\begin{eqnarray}
&&
(i\frac{\partial}{\partial t}-\varepsilon_\alpha)
G^{(i{\rm p}j{\rm h})}_{\alpha;\alpha'}(t-t')
\nonumber \\
&=&
\delta(t-t'){\cal N}_{\alpha;\alpha'}
+
\sum_{\beta}M^0_{\alpha;\beta}G^{(i{\rm p}j{\rm h})}_{\beta;\alpha'}(t-t')
\nonumber \\
&+&
\sum_{\beta}\int dt''
M_{\alpha;\beta}(t-t'')G^{(i{\rm p}j{\rm h})}_{\beta;\alpha'}(t''-t'),
\label{eq-ipjhGFdyson}
\end{eqnarray}
with
\begin{eqnarray}
&&
M^0_{\alpha;\alpha'}
=
\sum_{\beta}\langle  [[A_{\alpha},V]_{-},A_{\beta}^\dagger]_{\pm} \rangle
{\cal N}_{\beta;\alpha'}^{-1}
\label{eq-staticmass}\\
&&
M_{\alpha;\alpha'}(t-t')
\nonumber \\
&=&
-i\sum_{\beta}
\langle T([A_{\alpha}(t),V]_{-}[V,A_{\beta}^\dagger(t')]_{-})\rangle_{\rm irr.}
{\cal N}_{\beta;\alpha'}^{-1},
\label{eq-dynamicalmass}\\
&&
{\cal N}_{\alpha;\alpha'}
=
\langle [A_{\alpha},A_{\alpha'}^\dagger]_{\pm}\rangle
\end{eqnarray}
where 
$\varepsilon_\alpha$ is defined by 
$[A_{\alpha},T-\mu N]_{-}=\varepsilon_\alpha A_{\alpha}$, 
and ${\cal N}_{\alpha;\alpha'}^{-1}$ is the inverse of the matrix 
${\cal N}_{\alpha;\alpha'}$.

According to the time dependence, 
we shall call $M^0_{\alpha;\alpha'}$ static mass operator
and $M_{\alpha;\alpha'}(t-t')$ dynamical mass operator.

Since $[A_{\alpha}(t),V]_{-}$ is the operator $A_{\alpha}$
augmented by one annihilation and one creation operator,
$\langle T([A_{\alpha}(t),V]_{-}[V,A_{\beta}^\dagger(t'')]_{-})\rangle_{\rm irr.}$
becomes a ($i+1$)p($j+1$)h GF in the dynamical mass of 
Eq.~(\ref{eq-dynamicalmass}).
The index `irr.' in Eq.~(\ref{eq-dynamicalmass}) 
stands for the mass operator being irreducible 
with respect to a cut of $i$p$j$h lines~\cite{rs}.

From the Dyson equation of Eq.~(\ref{eq-ipjhGFdyson})
we can see that 
the Fourier transform of the bare $i$p$j$h GF is 
\begin{eqnarray}
G^{0(i{\rm p}j{\rm h})}_{\alpha;\alpha'}(\omega)
=\frac{{\cal N}_{\alpha;\alpha'}}
      {\omega-\varepsilon_{\alpha}}.
\label{eq-bareipjhGF}
\end{eqnarray}

\section{\label{sec-appendix-alphamass}Quartet mass operator}

Here we derive the quartet  mass operator 
of Eq.~(\ref{eq-fullalphaBECmass})
shown in Sec.~\ref{sec-quartet}.

Notice that we shall use summation convention for repeated indices
and neglect all terms except the ones which are associated with 
the quartet order parameters.

From Eq.~(\ref{eq-ipjhGFdyson}),
the Dyson equation for the $1$p GF is~\cite{rs}
\begin{eqnarray}
&&
\left(
i\frac{\partial }{\partial t}
-\varepsilon_1\right)
G^{(1{\rm p})}_{1;2}(t_1-t_2)
\nonumber \\
&=&
\delta_{12}\delta(t_1-t_2)
\nonumber \\
&+&
\int dt_3 M^{(1{\rm p})}_{1;3}(t_1-t_3)G^{(1{\rm p})}_{3;2}(t_3-t_2).
\end{eqnarray}

The Fourier transform yields
\begin{eqnarray}
(\omega-\varepsilon_1)G^{(1{\rm p})}_{1;2}(\omega)
=\delta_{12}
+M^{(1{\rm p})}_{1;3}(\omega)G^{(1{\rm p})}_{3;2}(\omega)
\end{eqnarray}
with 
\begin{eqnarray}
M^{(1{\rm p})}_{1;2}(\omega)
=
\frac{1}{2}\bar v_{1z_1,a_1a_1'}
G^{(2{\rm p}1{\rm h}){\rm irr.}}_{a_1a_1',z_1;a_2a_2',z_2}(\omega) 
\frac{1}{2}\bar v_{2z_2,a_2a_2'}.
\label{eq-mass-1p-2p1h}
\end{eqnarray}

\noindent

Therefore
\begin{eqnarray}
&&
G^{(1{\rm p})}_{1;2}(\omega)
\nonumber \\
&=&
G^{0(1{\rm p})}_{1;2}(\omega)
+G^{0(1{\rm p})}_{1;3}(\omega)
M^{(1{\rm p})}_{3;4}(\omega)
G^{(1{\rm p})}_{4;2}(\omega),
\end{eqnarray}
where
\begin{eqnarray}
G^{0(1{\rm p})}_{1;2}(\omega)
=\frac{\delta_{12}}{\omega-\varepsilon_1}.
\label{eq-free1pGF}
\end{eqnarray}

For the Dyson equation of the $2$p1h GF one obtains
\begin{eqnarray}
&&
\left(\omega-(\varepsilon_{a_1}+\varepsilon_{a_1'}-\varepsilon_{z_1})\right)
G^{(2{\rm p}1{\rm h})}_{a_1a_1',z_1;a_2a_2',z_2}(\omega)
\nonumber \\
&=&
{\cal N}_{a_1a_1',z_1;a_2a_2',z_2}
\nonumber \\
&+&
M^{(2{\rm p}1{\rm h})}_{a_1a_1',z_1;a_3a_3',z_3}(\omega)
G^{(2{\rm p}1{\rm h})}_{a_3a_3',z_3;a_2a_2',z_2}(\omega),
\end{eqnarray}
where
\begin{eqnarray}
&&
M^{(2{\rm p}1{\rm h})}_{a_1a_1',z_1;a_2a_2',z_2}(\omega)
\nonumber \\
&=&
\Bigl[
 \frac{1}{2}\bar v_{a_1y_1,b_1b_1'}\delta_{z_1y_1'}\delta_{a_1'b_1''}
+\frac{1}{2}\bar v_{a_1'y_1',b_1b_1'}\delta_{z_1y_1}\delta_{a_1b_1''}
\nonumber \\
&&
-\frac{1}{2}\bar v_{z_1b_1'',y_1y_1'}\delta_{a_1b_1}\delta_{a_1'b_1'}
\Bigr]
G^{(3{\rm p}2{\rm h}){\rm irr.}}_{b_1b_1'b_1'',y_1y_1';b_2b_2'b_2'',y_2y_2'}
(\omega)
\nonumber \\
&\times&
\Bigl[
 \frac{1}{2}\bar v_{a_3y_2,b_2b_2'}\delta_{z_3y_2'}\delta_{a_3'b_2''}
+\frac{1}{2}\bar v_{a_3'y_2',b_2b_2'}\delta_{z_3y_2}\delta_{a_3b_3''}
\nonumber \\
&&
-\frac{1}{2}\bar v_{z_3b_2'',y_2y_2'}\delta_{a_3b_2}\delta_{a_3'b_2'}
\Bigr]
{\cal N}_{a_3a_3',z_3;a_2a_2',z_2}^{-1}
\end{eqnarray}
and
\begin{eqnarray}
{\cal N}_{a_1a_1',z_1;a_2a_2',z_2}
=\langle [c_{z_1}^\dagger c_{a_1'} c_{a_1}, 
          c_{a_2}^\dagger c_{a_2'}^\dagger c_{z_2}]_{+}\rangle.
\end{eqnarray}

Therefore
\begin{eqnarray}
&&
G^{(2{\rm p}1{\rm h})}_{a_1a_1',z_1;a_2a_2',z_2}(\omega)
\nonumber \\
&=&
G^{0(2{\rm p}1{\rm h})}_{a_1a_1',z_1;a_2a_2',z_2}(\omega)
\nonumber \\
&+&
G^{0(2{\rm p}1{\rm h})}_{a_1a_1',z_1;a_3a_3',z_3}(\omega)
{\cal N}_{a_3a_3',z_3;a_4a_4',z_4}^{-1}
\nonumber \\
&&\times
M^{(2{\rm p}1{\rm h})}_{a_4a_4',z_4;a_5a_5',z_5}(\omega)
G^{(2{\rm p}1{\rm h})}_{a_5a_5',z_5;a_2a_2',z_2}(\omega),
\label{eq-2p1hGF}
\end{eqnarray}
with
\begin{eqnarray}
G^{0(2{\rm p}1{\rm h})}_{a_1a_1',z_1;a_2a_2',z_2}(\omega)
=\frac{{\cal N}_{a_1a_1',z_1;a_2a_2',z_2}}
      {\omega-(\varepsilon_{a_1a_1'}-\varepsilon_{z_1})}
\label{eq-free2p1hGF}
\end{eqnarray}
and 
\begin{eqnarray}
{\cal N}_{a_1a_1',z_1;a_3a_3',z_3}
{\cal N}_{a_3a_3',z_3;a_2a_2',z_2}^{-1}
=\delta_{a_1a_2}\delta_{a_1'a_2'}\delta_{z_1z_2},
\label{eq-inverse2p1hN}
\end{eqnarray}
where $\varepsilon_{ij...}=\varepsilon_{i}+\varepsilon_{j}+\cdots$
in (\ref{eq-free2p1hGF}).

Substituting the 2p1h GF of Eq.~(\ref{eq-2p1hGF})
into the mass operator of Eq.~(\ref{eq-mass-1p-2p1h}) 
we obtain
\begin{eqnarray}
&&
M^{(1{\rm p})}_{1;2}(\omega)
\nonumber \\
&=&\Gamma^{(3)}_{1y_1y_1',b_1b_1'b_1''}(\omega)
G^{(3{\rm p}2{\rm h}){\rm irr.}}_{b_1b_1'b_1'',y_1y_1';b_2b_2'b_2'',y_2y_2'}
(\omega)
\nonumber \\
&&\times
\Gamma^{(3)*}_{2y_2y_2',b_2b_2'b_2''}(\omega),
\label{eq-mass-1p-3p2h}
\end{eqnarray}
where
\begin{eqnarray}
&&
\Gamma^{(3)}_{1y_1y_1',b_1b_1'b_1''}(\omega)
\nonumber \\
&=&\frac{1}{2}\bar v_{1z_1,a_1a_1'}
G^{0(2{\rm p}1{\rm h})}_{a_1a_1',z_1;a_3a_3',z_3}(\omega)
{\cal N}_{a_3a_3',z_3;a_4a_4',z_4}^{-1}
\nonumber \\
&\times&
\Bigl[
 \frac{1}{2}\bar v_{a_4y_1,b_1b_1'}\delta_{z_1y_1'}\delta_{a_4'b_1''}
+\frac{1}{2}\bar v_{a_4'y_1',b_1b_1'}\delta_{z_1y_1}\delta_{a_4b_1''}
\nonumber \\
&&-\frac{1}{2}\bar v_{z_4b_1'',y_1y_1'}\delta_{a_4b_1}\delta_{a_4'b_1'}
\Bigr].
\label{eq-gamma3}
\end{eqnarray}
In (\ref{eq-mass-1p-3p2h})
we omitted the term derived from the first term at r.h.s. of (\ref{eq-2p1hGF}) 
because it is disconnected with  the quartet order parameter.
In the effective three body vertex $\Gamma^{(3)}$
of Eq.~(\ref{eq-gamma3})
we give not the exact 2p1h GF but the free one,
as we only consider in this work the lowest order approximation, 
though the exact 2p1h GF figures, in principle, in the right vertex 
$\Gamma^{(3)}$ of 
Eq.~(\ref{eq-gamma3}) substituting (\ref{eq-2p1hGF}) 
into (\ref{eq-mass-1p-2p1h}).
In Eq.~(\ref{eq-gamma3}), 
using Eqs.~(\ref{eq-free2p1hGF}) and (\ref{eq-inverse2p1hN}), we obtain
\begin{eqnarray}
&&
G^{0(2{\rm p}1{\rm h})}_{a_1a_1',z_1;a_3a_3',z_3}(\omega)
{\cal N}_{a_3a_3',z_3;a_4a_4',z_4}^{-1}
\nonumber \\
&=&
\frac{\delta_{a_1a_4}\delta_{a_1'a_4'}\delta_{z_1z_4}}
      {\omega-\varepsilon_{a_1a_1'}+\varepsilon_{z_1}},
\end{eqnarray}
and thus Eq.~(\ref{eq-gamma3}) 
is consistent with Eq.~(\ref{eq-Gamma3}).

\begin{figure}
\includegraphics[width=70mm]{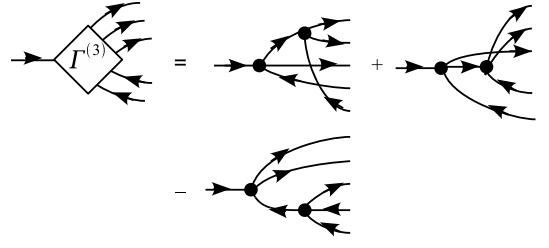}
\caption{\label{fig-gamma3}Graphical representation of $\Gamma^{(3)}$
in Eq.~(\ref{eq-gamma3}).
The dots is represent the two-body interaction.}
\end{figure}

Furthermore, the Dyson equation for 3p2h GF is
\begin{eqnarray}
&&(\omega-(\varepsilon_{b_1b_1'b_1''}-\varepsilon_{y_1y_1'}))
G^{(3{\rm p}2{\rm h})}_{b_1b_1'b_1'',y_1y_1';b_2b_2'b_2'',y_2y_2'}(\omega)
\nonumber \\
&=&{\cal N}_{b_1b_1'b_1'',y_1y_1';b_2b_2'b_2'',y_2y_2'}
\nonumber \\
&+&
M^{(3{\rm p}2{\rm h})}_{b_1b_1'b_1'',y_1y_1';b_3b_3'b_3'',y_3y_3'}(\omega)
\nonumber \\
&&\times
G^{(3{\rm p}2{\rm h})}_{b_3b_3'b_3'',y_3y_3';b_2b_2'b_2'',y_2y_2'}(\omega).
\end{eqnarray}
The mass operator is then given by
\begin{eqnarray}
&&M^{(3{\rm p}2{\rm h})}_{b_1b_1'b_1'',y_1y_1';b_2b_2'b_2'',y_2y_2'}(\omega)
\nonumber \\
&=&
\Bigl[
 \frac{1}{2}\bar v_{b_1x_1'',c_1c_1'''}
 \delta_{y_1x_1}\delta_{y_1'x_1'}\delta_{b_1'c_1'}\delta_{b_1''c_1''}
\nonumber \\
&&
+\frac{1}{2}\bar v_{b_1'x_1'',c_1'c_1'''}
\delta_{y_1x_1}\delta_{y_1'x_1'}\delta_{b_1c_1}\delta_{b_1''c_1''}
\nonumber \\
&&
+\frac{1}{2}\bar v_{b_1''x_1'',c_1''c_1'''}
\delta_{y_1x_1}\delta_{y_1'x_1'}\delta_{b_1c_1}\delta_{b_1'c_1'}
\nonumber \\
&&
-\frac{1}{2}\bar v_{y_1c_1''',x_1x_1''}
\delta_{y_1'x_1'}\delta_{b_1c_1}\delta_{b_1'c_1'}\delta_{b_1''c_1''}
\nonumber \\
&&
-\frac{1}{2}\bar v_{y_1'c_1''',x_1'x_1''}
\delta_{y_1x_1}\delta_{b_1c_1}\delta_{b_1'c_1'}\delta_{b_1''c_1''}
\Bigr]
\nonumber \\
&&\times
G^{(4{\rm p}3{\rm h}){\rm irr.}}_{c_1c_1'c_1''c_1''',x_1x_1'x_1'';c_2c_2'c_2''c_2''',x_2x_2'x_2''}
(\omega)
\nonumber \\
&\times&
\Bigl[
 \frac{1}{2}\bar v_{b_3x_2'',c_2c_2'''}
 \delta_{y_3x_2}\delta_{y_3'x_2'}\delta_{b_3'c_2'}\delta_{b_3''c_2''}
\nonumber \\
&&
+\frac{1}{2}\bar v_{b_3'x_2'',c_2'c_2'''}
\delta_{y_3x_2}\delta_{y_3'x_2'}\delta_{b_3c_2}\delta_{b_3''c_2''}
\nonumber \\
&&
+\frac{1}{2}\bar v_{b_3''x_2'',c_2''c_2'''}
\delta_{y_3x_2}\delta_{y_3'x_2'}\delta_{b_3c_2}\delta_{b_3'c_2'}
\nonumber \\
&&
-\frac{1}{2}\bar v_{y_3c_2''',x_2x_2''}
\delta_{y_3'x_2'}\delta_{b_3c_2}\delta_{b_3'c_2'}\delta_{b_3''c_2''}
\nonumber \\
&&
-\frac{1}{2}\bar v_{y_3'c_2''',x_2'x_2''}
\delta_{y_3x_2}\delta_{b_3c_2}\delta_{b_3'c_2'}\delta_{b_3''c_2''}
\Bigr]
\nonumber \\
&&\times
{\cal N}_{b_3b_3'b_3'',y_3y_3';b_2b_2'b_2'',y_2y_2'}^{-1},
\end{eqnarray}
and
\begin{eqnarray}
&&
{\cal N}_{b_1b_1'b_1'',y_1y_1';b_2b_2'b_2'',y_2y_2'}
\nonumber \\
&=&
\langle
 [c_{y_1}^\dagger c_{y_1'}^\dagger c_{b_1''} c_{b_1'} c_{b_1},
  c_{b_1}^\dagger c_{b_1'}^\dagger  c_{b_1''}^\dagger c_{y_1'}c_{y_1}]_{+}
 \rangle.
\end{eqnarray}

Therefore, one obtains for the 3p2h GF
\begin{eqnarray}
&&G^{(3{\rm p}2{\rm h})}_{b_1b_1'b_1'',y_1y_1';b_2b_2'b_2'',y_2y_2'}(\omega)
\nonumber \\
&=&
G^{0(3{\rm p}2{\rm h})}_{b_1b_1'b_1'',y_1y_1';b_2b_2'b_2'',y_2y_2'}(\omega)
\nonumber \\
&+&
G^{0(3{\rm p}2{\rm h})}_{b_1b_1'b_1'',y_1y_1';b_3b_3'b_3'',y_3y_3'}(\omega)
{\cal N}_{b_3b_3'b_3'',y_3y_3';b_4b_4'b_4'',y_4y_4'}^{-1}
\nonumber \\
&&\times
M^{(3{\rm p}2{\rm h})}_{b_4b_4'b_4'',y_4y_4';b_5b_5'b_5'',y_5y_5'}(\omega)
\nonumber \\
&&\times
G^{(3{\rm p}2{\rm h})}_{b_5b_5'b_5'',y_5y_5';b_2b_2'b_2'',y_2y_2'}(\omega),
\label{eq-3p2hGF}
\end{eqnarray}
where
\begin{eqnarray}
&&
G^{0(3{\rm p}2{\rm h})}_{b_1b_1'b_1'',y_1y_1';b_2b_2'b_2'',y_2y_2'}(\omega)
\nonumber \\
&=&
\frac{{\cal N}_{b_1b_1'b_1'',y_1y_1';b_2b_2'b_2'',y_2y_2'}}
      {\omega-(\varepsilon_{b_1b_1'b_1''}-\varepsilon_{y_1y_1'})}.
\end{eqnarray}

\begin{figure}
\includegraphics[width=72mm]{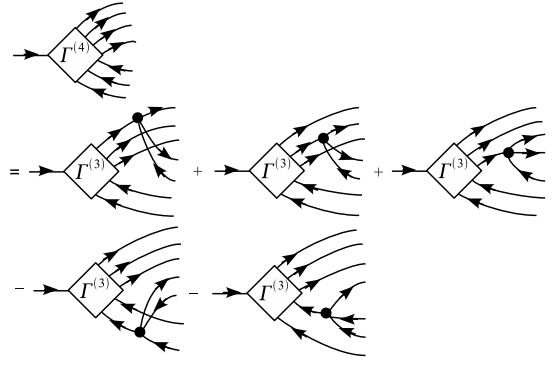}
\caption{\label{fig-gamma4}Graphical representation of $\Gamma^{(4)}$
in Eq.~(\ref{eq-gamma4}).}
\end{figure}

Substituting 3p2h GF of (\ref{eq-3p2hGF}) into the mass operator 
of Eq.~(\ref{eq-mass-1p-3p2h}) leads to
\begin{eqnarray}
&&
M^{(1{\rm p})}_{1;2}(\omega)
\nonumber \\
&=&
\Gamma^{(4)}_{1x_1x_1'x_1'',c_1c_1'c_1''c_1'''}(\omega)
\nonumber \\
&&\times
G^{(4{\rm p}3{\rm h}){\rm irr.}}_{c_1c_1'c_1''c_1''',x_1x_1'x_1'';c_2c_2'c_2''c_2''',x_2x_2'x_2''}
(\omega)
\nonumber \\
&&\times
\Gamma^{(4)*}_{2x_2x_2'x_2'',c_2c_2'c_2''c_2'''}(\omega),
\label{eq-quartetmassoperator}
\end{eqnarray}
with
\begin{eqnarray}
&&
\Gamma^{(4)}_{1x_1x_1'x_1'',c_1c_1'c_1''c_1'''}(\omega)
\nonumber \\
&=&\Gamma^{(3)}_{1y_1y_1',b_1b_1'b_1''}(\omega)
\nonumber \\
&&\times
G^{0(3{\rm p}2{\rm h})}_{b_1b_1'b_1'',y_1y_1';b_3b_3'b_3'',y_3y_3'}(\omega)
{\cal N}_{b_3b_3'b_3'',y_3y_3';b_4b_4'b_4'',y_4y_4'}^{-1}
\nonumber \\
&\times&
\Bigl[
 \frac{1}{2}\bar v_{b_4x_1'',c_1c_1'''}
 \delta_{y_4x_1}\delta_{y_4'x_1'}\delta_{b_4'c_1'}\delta_{b_4''c_1''}
\nonumber \\
&&
+\frac{1}{2}\bar v_{b_4'x_1'',c_1'c_1'''}
\delta_{y_4x_1}\delta_{y_4'x_1'}\delta_{b_4c_1}\delta_{b_4''c_1''}
\nonumber \\
&&
+\frac{1}{2}\bar v_{b_4''x_1'',c_1''c_1'''}
\delta_{y_4x_1}\delta_{y_4'x_1'}\delta_{b_4c_1}\delta_{b_4'c_1'}
\nonumber \\
&&
-\frac{1}{2}\bar v_{y_4c_1''',x_1x_1''}
\delta_{y_4'x_1'}\delta_{b_4c_1}\delta_{b_4'c_1'}\delta_{b_4''c_1''}
\nonumber \\
&&
-\frac{1}{2}\bar v_{y_4'c_1''',x_1'x_1''}
\delta_{y_4x_1}\delta_{b_4c_1}\delta_{b_4'c_1'}\delta_{b_4''c_1''}
\Bigr]
\label{eq-gamma4}
\end{eqnarray}
Here we also omitted the terms associated with the first term 
at r.h.s. of Eq.~(\ref{eq-3p2hGF}) because it is disconnected 
with  the quartet order parameter. 
Besides we introduced the free 3p2h GF in $\Gamma^{(4)}$
of (\ref{eq-gamma4})
for the same reason as this was done in Eqs.~(\ref{eq-mass-1p-3p2h}) 
and (\ref{eq-gamma3}).
The graphical representation of $\Gamma^{(4)}$ is shown 
in Fig.~\ref{fig-gamma4}.

The 4p3h GF can, for our purpose of $\alpha$-particle condensation, 
approximately be decomposed into the order parameter and 
the free three body GF as follows (in analogy to what is done in the 
pairing case)
\begin{eqnarray}
&&
G^{(4{\rm p}3{\rm h})}
_{c_1c_1'c_1''c_1''',x_1x_1'x_1'';c_2c_2'c_2''c_2''',x_2x_2'x_2''}
(t_1-t_2)
\nonumber \\
&=&
-i
\langle T
(c^\dagger_{x_1}c^\dagger_{x_1'}c^\dagger_{x_1''}
c_{c_1'''}c_{c_1''}c_{c_1'}c_{c_1})_{t_1}
(c^\dagger_{c_2}c^\dagger_{c_2'}c^\dagger_{c_2''}c^\dagger_{c_2'''}
c_{x_2''}c_{x_2'}c_{x_2})_{t_2}
\rangle
\nonumber \\
&=&
-i
\langle T
(c_{c_1'''}c_{c_1''}c_{c_1'}c_{c_1})_{t_1}
\rangle
\langle T
(c^\dagger_{c_2}c^\dagger_{c_2'}c^\dagger_{c_2''}c^\dagger_{c_2'''})_{t_2}
\rangle
\nonumber \\
&&\times
\langle T
(c^\dagger_{x_1}c^\dagger_{x_1'}c^\dagger_{x_1''})_{t_1}
(c_{x_2''}c_{x_2'}c_{x_2})_{t_2}
\rangle
\nonumber \\
&=&
\langle 
c_{c_1'''}c_{c_1''}c_{c_1'}c_{c_1} 
\rangle
\langle 
c^\dagger_{c_2}c^\dagger_{c_2'}c^\dagger_{c_2''}c^\dagger_{c_2'''}
\rangle
\nonumber \\
&&\times
G^{0(3{\rm h})}_{x_1x_1'x_1'';x_2x_2'x_2''}(t_1-t_2)
\end{eqnarray}

The Fourier transform  is 
\begin{eqnarray}
&&
G^{(4{\rm p}3{\rm h})}
_{c_1c_1'c_1''c_1''',x_1x_1'x_1'';c_2c_2'c_2''c_2''',x_2x_2'x_2''}
(\omega)
\nonumber \\
&=&
\langle 
c_{c_1'''}c_{c_1''}c_{c_1'}c_{c_1} 
\rangle
\langle 
c^\dagger_{c_2}c^\dagger_{c_2'}c^\dagger_{c_2''}c^\dagger_{c_2'''}
\rangle
\nonumber \\
&&\times
G^{0(3{\rm h})}_{x_1x_1'x_1'';x_2x_2'x_2''}(\omega).
\label{eq-fourier4p3hGF}
\end{eqnarray}

The free 3h GF 
is explicitly given  by
\begin{eqnarray}
&&
G^{0(3{\rm h})}_{x_1x_1'x_1'';x_2x_2'x_2''}(\omega)
\nonumber \\
&=&
\frac{
\bar f_{x_1}
\bar f_{x_1'}
\bar f_{x_1''}
+
f_{x_1}
f_{x_1'}
f_{x_1''}}
{\omega+\varepsilon_{x_1x_1'x_1''}}
P_{x_1x_1'x_1'';x_2x_2'x_2''}
\end{eqnarray}
with $P_{x_1x_1'x_1'';x_2x_2'x_2''}$ as in Eq.~(\ref{eq-antisymmetric}).

Therefore, Eq.~(\ref{eq-fourier4p3hGF}) becomes
\begin{eqnarray}
&&
G^{(4{\rm p}3{\rm h})}
_{c_1c_1'c_1''c_1''',x_1x_1'x_1'';c_2c_2'c_2''c_2''',x_2x_2'x_2''}
(\omega)
\nonumber \\
&=&
\langle  c_{c_1'''}c_{c_1''}c_{c_1'}c_{c_1} \rangle
 \langle c_{c_2}^\dagger c_{c_2'}^\dagger 
                    c_{c_2''}^\dagger c_{c_2'''}^\dagger \rangle
\nonumber \\
&&\times
\frac{\bar f_{x_1}\bar f_{x_1'} \bar f_{x_1''}
     +f_{x_1} f_{x_1'} f_{x_1''}}
{\omega+\varepsilon_{x_1x_1'x_1''}}
P_{x_1x_1'x_1'';x_2x_2'x_2''}.
\end{eqnarray}

Finally we obtain for the quartet mass operator,
\begin{eqnarray}
&&
M^{\rm quartet}_{1;2}(\omega)
\nonumber \\
&=&
\Gamma^{(4)}_{1x_1x_1'x_1'';c_1c_1'c_1''c_1'''}(\omega)
\langle  c_{c_1'''}c_{c_1''}c_{c_1'}c_{c_1} \rangle
\nonumber \\
&&\times
\frac{\bar f_{x_1}\bar f_{x_1'} \bar f_{x_1''}
     +f_{x_1} f_{x_1'} f_{x_1''}}
{\omega+\varepsilon_{x_1x_1'x_1''}}
P_{x_1x_1'x_1'';x_2x_2'x_2''}
\nonumber \\
&&\times
 \langle c_{c_2}^\dagger c_{c_2'}^\dagger 
                    c_{c_2''}^\dagger c_{c_2'''}^\dagger \rangle
\Gamma^{(4)*}_{2x_2x_2'x_2'';c_2c_2'c_2''c_2'''}(\omega).
\label{eq-app-qmass}
\end{eqnarray}

\section{\label{sec-appendix-inmedium}in-medium quartet order 
parameter equation}

We here give a short derivation of  the in-medium quartet order parameter 
equation. It 
is derived from the Dyson equation for 4p GF 
in static approximation of the mass operator.
The Dyson equation for 4p GF derived from Eq.~(\ref{eq-ipjhGFdyson})
without dynamical mass operator is
\begin{eqnarray}
&&
\left(i\frac{\partial}{\partial t}-\varepsilon_{1234}\right)
G^{(4{\rm p})}_{1234;1'2'3'4'}(t-t')
\nonumber \\
&=&
{\cal N}_{1234;1'2'3'4'}
\nonumber \\
&+&
V_{1234;1''2''3''4''}G^{(4{\rm p})}_{1234;1'2'3'4'}(t-t')
\end{eqnarray}
with in the ladder approximation
\begin{eqnarray}
&&
V_{1234;1'2'3'4'}
\nonumber \\
&=&\langle [[c_{4}c_{3}c_{2}c_{1},V]_{-},
         c_{1''}^\dagger c_{2''}^\dagger c_{3''}^\dagger c_{4''}^\dagger ]_{-} 
 \rangle
{\cal N}_{1''2''3''4'';1'2'3'4'}^{-1}
\nonumber \\
&=&
 \frac{1}{2}(1-\rho_1-\rho_2)\bar v_{12,1'2'}\delta_{33'}\delta_{44'}
\nonumber \\
&+&
\frac{1}{2}(1-\rho_1-\rho_3)\bar v_{13,1'3'}\delta_{22'}\delta_{44'}
\nonumber \\
&+&
\frac{1}{2}(1-\rho_1-\rho_4)\bar v_{14,1'4'}\delta_{22'}\delta_{33'}
\nonumber \\
&+&
{\rm permutations}
\label{eq-inmediumfbint}
\end{eqnarray}
where we approximated the correlation functions by factorizing them 
into products of single particle occupation numbers.

The Fourier transform of 4p GF is
\begin{eqnarray}
&&
G^{(4{\rm p})}_{1234;1'2'3'4'}(\omega)
\nonumber \\
&=&
G^{0(4{\rm p})}_{1234;1'2'3'4'}(\omega)
\nonumber \\
&+&
G^{0(4{\rm p})}_{1234;1''2''3''4''}(\omega)
{\cal N}_{1''2''3''4'';5678}^{-1}
\nonumber \\
&&\times
V_{5678;5'6'7'8'}
G^{(4{\rm p})}_{5'6'7'8';1'2'3'4'}(\omega),
\label{eq-app-4bGF}
\end{eqnarray}
where 
\begin{eqnarray}
G^{0(4{\rm p})}_{1234;1'2'3'4'}(\omega)
=\frac{{\cal N}_{1234;1'2'3'4'}}{\omega-\varepsilon_{1234}}.
\end{eqnarray}

From the spectral representation of the 4p GF we shall only retain the 
ground state because of its condensate character.
Therefore, at the ground state pole, i.e. at $\omega=0$, we obtain
\begin{eqnarray}
\langle c_{4}c_{3}c_{2}c_{1}\rangle
=-\frac{1}{\varepsilon_{1234}}
V_{1234;1'2'3'4'}\langle c_{4'}c_{3'}c_{2'}c_{1'}\rangle.
\label{eq-app-inmediumfourbody}
\end{eqnarray}

This equation corresponds to Eq.~(\ref{eq-gapeq}) in the pairing case.

\section{\label{sec-appendix-discussion}
Approximate mass operator 
in Eq.~(\ref{eq-Qmassoperator})}

After the derivation of a single particle mass operator 
containing the quartet condensate in App.~\ref{sec-appendix-alphamass},
we easily recognize that its expression, for instance 
the vertex $\Gamma^{(4)}$ in Eqs.~(\ref{eq-Gamma4}) and (\ref{eq-Gamma3}) 
(or (\ref{eq-gamma3}) and (\ref{eq-gamma4})) it contains, is of 
considerable complexity, prohibitive for a direct numerical application,
especially due to high dimensional integrals.
We, therefore, will have to proceed to a quite intensive study
of vertices, followed by reasonable approximations, in order to 
reduce drastically the numerical difficulty of the expressions.

A first purely formal simplification which we will introduce, is to consider 
instead of the quartet case, only the trion case. Trions are fermions and one 
would have to develop a whole proper philosophy to introduce an order 
parameter for trions. However, we will ignore this difficulty here, proceed 
formally
as if a trion order parameter existed in the same way as a quartet order 
parameter, and explain for this much simpler case our strategy. 
This can be done 
without loss of generality and in the end we simply will give our 
results for the quartet case which can be derived in complete analogy to the 
trion case.

Let us, therefore, begin with the expression 
of the single particle mass operator containing a `trion condensate'
which analogously to Eq.~(\ref{eq-app-qmass}) is given by
\begin{eqnarray}
&&
M_{1;1'}(\omega)
\nonumber \\
&=&
\Gamma^{(3)}_{123;456}(\omega)\langle c_6c_5c_4 \rangle
\frac{\bar f_2\bar f_3-f_2f_3}{\omega+\varepsilon_{23}}
(\delta_{22'}\delta_{33'}-\delta_{23'}\delta_{32'})
\nonumber \\
&&\times
\langle c^\dagger_{4'}c^\dagger_{5'}c^\dagger_{6'} \rangle
\Gamma^{(3)*}_{1'2'3';4'5'6'}(\omega)
\label{eq-trionmass}
\end{eqnarray}
with $\Gamma^{(3)}$ the three body vertex already given 
in (\ref{eq-Gamma3}) and derived 
in Eq.~(\ref{eq-gamma3}).
The first and rather obvious approximation we perform is 
to make $\Gamma^{(3)}$ `instantaneous', that is $\omega$-independent.
A standard procedure for this is to put the vertex `on the energy shell'.
This procedure is not always defined unambiguously. 
In the present case, one possibility certainly is to put 
in $\Gamma^{(3)}$, $\omega=-\varepsilon_{23}$, i.e. the energy where the 
mass operator (\ref{eq-trionmass}) is resonant.

\begin{figure}
\includegraphics[width=50mm]{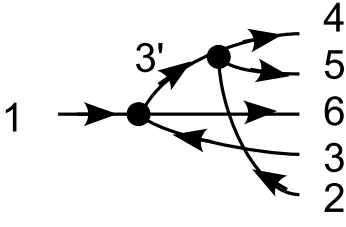}
\caption{\label{fig-firsttermofgamma3}
The graphical expression of 
the first term on the r.h.s. of (\ref{eq-Gamma3}).}
\end{figure}

Let us for the moment only investigate the first term on the r.h.s. of 
(\ref{eq-Gamma3}) with this prescription. We obtain
\begin{eqnarray}
&&
\Gamma^{(3,1)}_{123;456}(\omega=-\varepsilon_{23})
\nonumber \\
&&
=
\frac{1}{4}\bar v_{13,3'6}
\frac{1}{-\varepsilon_{2}-\varepsilon_{3'}-\varepsilon_{6}}
\bar v_{3'2,45}
\end{eqnarray}
A graphical interpretation of this term is given 
in Fig.~\ref{fig-firsttermofgamma3}.

\begin{figure}
\includegraphics[width=50mm]{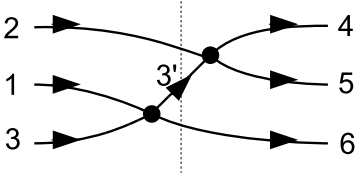}
\caption{\label{fig-figfirstG3threeGF}Topologically same diagram as in 
Fig.~\ref{fig-firsttermofgamma3}.}
\end{figure}

\noindent
This graph also can be interpreted as a particular second order term 
of a three body scattering process (123)$\to$(456), graphically 
represented in Fig.~\ref{fig-figfirstG3threeGF}.

\noindent
The intermediate propagator between the two vertices is given by
$\frac{1}{-\varepsilon_{2}-\varepsilon_{3'}-\varepsilon_{6}}$
(the energies are given by the propagators which are cut by the 
vertical line in Fig.~\ref{fig-figfirstG3threeGF}).
We, therefore, see that the static, on-shell part 
$\Gamma^{(3)}(\omega=-\varepsilon_{23})$ of (\ref{eq-trionmass})
can be interpreted as a second order three body scattering taken 
at $\omega=0$, i.e. in reality at three 
times the Fermi energy.
This certainly is a reasonable reduction of the second order process
to a static vertex $\Gamma^{(3,1)}$.

\begin{figure}
\includegraphics[width=50mm]{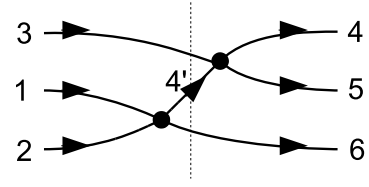}
\caption{\label{fig-figsecondG3threeGF}The graphical expression of 
the second term on the r.h.s. of (\ref{eq-Gamma3}).}
\end{figure}

Proceeding in the same way with the second term on r.h.s. 
of (\ref{eq-Gamma3}), we arrive at a second order 
process analogous to the one of Fig.~\ref{fig-figfirstG3threeGF}
as shown in Fig.~\ref{fig-figsecondG3threeGF}.
We see that Fig.~\ref{fig-figsecondG3threeGF} corresponds to 
Fig.~\ref{fig-figfirstG3threeGF} with particles 2 and 3 permuted.

\begin{figure}
\includegraphics[width=50mm]{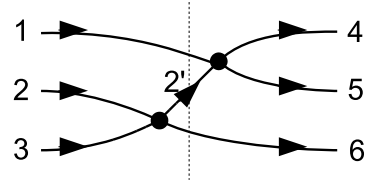}
\caption{\label{fig-figthirdG3threeGF}The graphical expression of 
the third term on the r.h.s. of (\ref{eq-Gamma3}).}
\end{figure}

Let us now consider the third term in (\ref{eq-Gamma3}).
In analogy to processes of Figs.~\ref{fig-figfirstG3threeGF}
and \ref{fig-figsecondG3threeGF} this corresponds to the graph of 
Fig.~\ref{fig-figthirdG3threeGF}.

\noindent
If we want to interpret this third term also as a static second order 
three body process with an intermediate propagator
$\frac{1}{-\varepsilon_{1}-\varepsilon_{2'}-\varepsilon_{6}}$
at $\omega=0$, contrary to first and second terms of (\ref{eq-Gamma3}),
we have to use as on shell prescription in (\ref{eq-Gamma3})
$\omega=\varepsilon_{1}+\varepsilon_{4}+\varepsilon_{5}+\varepsilon_{6}$. 
We, thus, suppose from now on that all vertices in $\Gamma^{(3)}$ are obtained 
from a second order three body scattering process at $\omega = 0$.

After these preliminary considerations,
let us now introduce in a phenomenological intuitive way a single particle
mass operator containing a trion order parameter
in a way analogous to the pairing case
\begin{equation}
M_{1;1'}(\omega)
=
\int \frac{d\omega'}{2\pi}
{\rm Im}T_{123;1'2'3'}(\omega-\omega')G^0_{23,2'3'}(-\omega'),
\label{eq-gerdmassop}
\end{equation}
with
\begin{eqnarray}
&&
{\rm Im}T_{123;1'2'3'}(\omega)
\nonumber \\
&=&
-2\pi\delta(\omega)\bar V_{123;456}
\psi^t_{456}\psi^{t *}_{4'5'6'}
\bar V_{4'5'6';1'2'3'},
\label{eq-3tm}
\end{eqnarray}
and
\begin{eqnarray}
&&
\bar V_{123;1'2'3'}
\nonumber \\
&=&
\frac{1}{2}
(\bar v_{12,1'2'}\delta_{33'}
+\bar v_{13,1'3'}\delta_{22'}
+\bar v_{23,2'3'}\delta_{11'}),
\label{eq-3bintnoPaulifactor}
\end{eqnarray}
where we abbreviated the the order parameter as
\begin{eqnarray}
\langle c_{3}c_{2}c_{1} \rangle
=\psi^t_{123}.
\end{eqnarray}

The `three body $T$-matrix' (\ref{eq-3tm}) in (\ref{eq-gerdmassop})
is contracted with the antisymmetrized two hole propagator 
\begin{eqnarray}
G^0_{23,2'3'}(-\omega)
=
-\frac{\bar f_2\bar f_3-f_2f_3}{\omega+\varepsilon_{23}}
(\delta_{22'}\delta_{33'}-\delta_{23'}\delta_{32'}).
\label{eq-twoholeGF}
\end{eqnarray}

With the above definitions (\ref{eq-gerdmassop})-(\ref{eq-twoholeGF})
we indeed see that the `trion' mass operator is constructed in 
full analogy to the one of the pairing case in Eq.~(\ref{eq-bcsmass}).
It is evident that in the quartet case we would define a four-body 
$T$-matrix as 
\begin{eqnarray}
{\rm Im}T_{1234;1'2'3'4'}(\omega)
=-2\pi\delta(\omega)\Delta_{1234}\Delta_{1'2'3'4'}^*
\end{eqnarray}
with 
\begin{eqnarray}
\Delta_{1234;5678}
=
\bar V_{1234;5678}
\langle c_{8}c_{7}c_{6}c_{5} \rangle
\end{eqnarray}
and 
\begin{eqnarray}
\bar V_{1234;1'2'3'4'}
&=&
\frac{1}{2}\bar v_{12;1'2'}\delta_{33'}\delta_{44'}
+
\frac{1}{2}\bar v_{13;1'3'}\delta_{22'}\delta_{44'}
\nonumber \\
&&
+{\rm permutations}
\label{eq-4bintnoPaulifactor}
\end{eqnarray}
and contract it with a 3h GF.

\begin{figure}
\includegraphics[width=50mm]{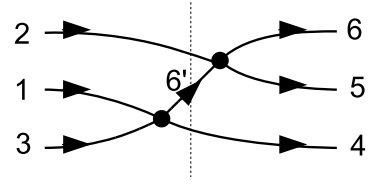}
\caption{\label{fig-figtildeG3third}The graphical expression of 
the third term on the r.h.s. of (\ref{eq-tildegamma3one}).}
\end{figure}

Let us now investigate whether we can make contact of the 
mass operator in (\ref{eq-gerdmassop}) with the one of (\ref{eq-trionmass}).
To this purpose, let us use in (\ref{eq-3tm}) the equation for 
the trion order parameter which in analogy to (\ref{eq-app-4bGF}) is given by 
\begin{eqnarray}
\psi_{123}^t
&=&
-\frac{1}{\varepsilon_{123}}
\Bigl[
(1-\rho_1-\rho_2)\frac{1}{2}\bar v_{12,1'2'}\delta_{33'}
\nonumber \\
&&
+(1-\rho_1-\rho_3)\frac{1}{2}\bar v_{13,1'3'}\delta_{22'}
\nonumber \\
&&
+(1-\rho_2-\rho_3)\frac{1}{2}\bar v_{23,2'3'}\delta_{11'}
\Bigr]
\psi_{1'2'3'}^t
\label{eq-inmediumthree}
\end{eqnarray}
and investigate for the moment only the expression corresponding to the 
second term on the r.h.s. of (\ref{eq-3bintnoPaulifactor}).
We obtain 
\begin{eqnarray}
&&
\frac{1}{2}\bar v_{13,46}\delta_{25}\psi^t_{456}
\nonumber \\
&=&
\frac{1}{2}\bar v_{13,46}\delta_{25}
\frac{1}{-\varepsilon_{123}}
\Bigl[
(1-\rho_4-\rho_5)\frac{1}{2}\bar v_{45,4'5'}\delta_{66'}
\nonumber \\
&&
+(1-\rho_4-\rho_6)\frac{1}{2}\bar v_{46,4'6'}\delta_{55'}
\nonumber \\
&&
+(1-\rho_5-\rho_6)\frac{1}{2}\bar v_{56,5'6'}\delta_{44'}
\Bigr]
\psi_{4'5'6'}^t.
\label{eq-3tmsecondorder}
\end{eqnarray}
We notice some similarity with the second order vertex 
(\ref{eq-trionmass}). The difference stems
from the presence of the occupation numbers $\rho_i$ 
in (\ref{eq-3tmsecondorder}) and from the fact that there is
more than one term. In principle the occupation numbers are 
the correlated ones. In the main part of the paper, we have 
seen that our theory practically only is valid for 
$\mu<0$ implying that the occupation numbers remain small, 
see Fig.~\ref{fig-spwf}. We, therefore, can neglect the 
occupation numbers in (\ref{eq-3tmsecondorder}) to good approximation 
(for the $\mu<0$ regime). Then, we can read off a vertex from 
(\ref{eq-3tmsecondorder})  
of the following form
\begin{eqnarray}
&&
\tilde \Gamma_{123;456}^{(3,2)}
\nonumber \\
&=&
\frac{1}{4}\bar v_{13,3'6}
\frac{1}{-\varepsilon_{23'6}}\bar v_{3'2,45}
+\frac{1}{4}\bar v_{13,3'6'}\delta_{25}
\frac{1}{-\varepsilon_{23'6'}}\bar v_{3'6',46}
\nonumber \\
&&
+\frac{1}{4}\bar v_{13,46'}
\frac{1}{-\varepsilon_{246'}}\bar v_{26',56}
\label{eq-tildegamma3one}
\end{eqnarray}
It is realized that the first term is equal to expression 
(\ref{eq-trionmass}). The second term is disconnected and,
thus, is an inproper term for a vertex (see discussion below). 
The third term can 
be graphically represented as shown in Fig.~\ref{fig-figtildeG3third}.
We see that the term of Fig.~\ref{fig-figtildeG3third} is 
obtained from Fig.~\ref{fig-figfirstG3threeGF} in permuting 
indices 4 and 6. It is thus equivalent to the first term 
in (\ref{eq-tildegamma3one}). This game can be repeated 
in the same way for the first and third term in (\ref{eq-3bintnoPaulifactor}).
The conclusion is always the same: there are twice as many terms
as equivalent terms from the on shell version of (\ref{eq-Gamma3}) and
in addition each time, there appears a disconnected term which 
should not be present in a vertex. This latter problem is, however, not a real 
one. The order parameter is fully correlated and no disconnected pieces can 
appear. The term which is disconnected in (\ref{eq-3tmsecondorder}) 
will certainly become 
connected in higher orders. Therefore such disconnected terms only serve to 
renormalize the vertices of the connected terms. In the end all bare vertices 
in the connected diagrams should be replaced by two body $T$-matrices and no 
disconnected terms would appear anymore. We will not further dwell on this 
point and simply discard the disconnected terms.

The conclusion, therefore, is that from (\ref{eq-gerdmassop}) 
we get twice as many terms for 
the vertex $\Gamma^{(3)}$ as is given in Eq.~(\ref{eq-Gamma3}). 
Otherwise the terms are 
equivalent under the condition that in (\ref{eq-Gamma3}) 
we take the above discussed 
on shell prescription. The correct procedure, therefore, is to divide the 
vertices $V_{123;1'2'3'}$ in (\ref{eq-3tm}) by a factor two.
In the quartet case we have to divide the vertex 
in (\ref{eq-4bintnoPaulifactor}) by a factor of 8.

In our numerical application (quartet case), 
instead of calculating all the 36 terms,
where several ones give equal contributions,  
resulting from squaring the vertex  (\ref{eq-4bintnoPaulifactor}) 
where some are more difficult to 
calculate than others, we take a very pragmatic point of view and keep only 
the first term on the r.h.s. of (\ref{eq-4bintnoPaulifactor}) 
simulating all the others by a 
factor $\lambda'$ as in Eq.~(\ref{eq-tildedelta}). 
In doing so, we suppose that all terms 
have more or less the same analytic structure. This is certainly the case, 
since all the terms are dominated by the behavior of the 3h level density 
whose typical structure is displayed in Sec.~\ref{sec-leveldensity}. 
It also is fortunate that 
the final result does not depend on $\lambda'$, as we now will demonstrate 
on the example of pairing.

\section{\label{sec-appendix-parameter}
Independence of  the parameter $\lambda'$
in Eq.~(\ref{eq-tildedelta})}

All arguments and derivations for the 
three-body vertices can directly be generalized to the four body case. 
In view of the fact that our favorite expression (\ref{eq-fullalphaBECmass}) 
is too complex for a direct numerical realization, we take a very pragmatic 
point of view for a pilot application. We have seen that some terms appear 
in all three forms we discussed for the vertices. We shall take one of those 
terms for numerical calculations with a parameter $\lambda'$ 
in front of the vertex which 
should mock up the influence of factors and also of topologically 
different graphs. However, we want to stress again that we do not think 
that topologically different vertices will finally give rise to different 
analytic structure of the mass operator. Essentially these additional terms 
will again only renormalize the vertices.
The form which we then retain is just the first one on the r.h.s. of (D10), 
since it is the one where the single particle 
motion directly couples via an interaction to the quartet amplitude, that is

\begin{equation}
\tilde \Delta_{1234} = 
\lambda'\frac{1}{2}\bar v_{12,1'2'}\delta_{33'}\delta_{44'}
\langle c_{1'}c_{2'}c_{3'}c_{4'} \rangle
\end{equation}

\noindent
With the separable form of the two body interaction~\cite{slr09},
we then 
obtain expression (\ref{eq-appqmo}).
As mentioned already, we will approximately absorb factors and  all the 
other terms in 
renormalizing the vertex by a constant factor $\lambda'$. This shall be a 
quite valid 
procedure, since the analytic structure of 
all the other terms is very similar.
Fortunately our results will not depend on the strength 
of the renormalization factor $\lambda'$ of the vertex. 
This statement may be surprising but can be explained at hand of the 
standard BCS example as follows.

Schematically, we write expression for single particle GF in BCS 
approximation of Eqs.~(\ref{eq-bcs1pGF}) and (\ref{eq-bcsmass})
as follows (with self evident notation)

\begin{equation}
G^{\rm BCS} = 
\frac{1}{\displaystyle \omega - \varepsilon-
\frac{v\langle cc \rangle 
\langle c^\dagger c^\dagger \rangle v}
{\omega+\varepsilon}}
\end{equation}

\noindent
Now let us change in above equation vertex $v$ to $v' = \lambda'v$. 
Let us apply to 
all quantities containing this new vertex a `prime'. 
Then BCS single particle GF is given by

\begin{equation}
{G'}^{\rm BCS} = \frac{1-\rho'}{\omega - E'} + \frac{\rho'}{\omega + E'}
\end{equation}

\noindent
with ${E'}^2 = \varepsilon^2 + {D'}^2$, $D' = v'\langle cc \rangle$,
and $\rho'=(1-\varepsilon/E')/2$.
The equation for order parameter then reads

\begin{equation}
\langle cc \rangle = -\frac{1-2\rho'}{2\varepsilon}v\langle cc \rangle
\end{equation}

\noindent
where we should pay attention that the $v$ in that equation is still the 
original one, as in our four-body order parameter equation 
of (\ref{eq-inmediumfbint}) and (\ref{eq-app-inmediumfourbody}).

From above we obtain the gap equation for $\Delta '$

\begin{equation}
\Delta ' = -\sum v\frac{\Delta '}{2E'}
\end{equation}

\noindent
Multiplying this equation with $\lambda'$, we see that $E'=E$ and, therefore, 
nothing is changed in multiplying in BCS single particle mass operator the 
vertex by an arbitrary factor. Translated to our quartet problem, this 
implies that the question of multiplicity of diagrams discussed above 
is of no consequence for our numerical results. Of course, the addition 
of topologically different contributions to the vertex $\Gamma^{(4)}$ can 
change things slightly but again one can imagine that to good approximation 
it only renormalizes the vertex corresponding to the process which we treat 
explicitly. Therefore, our 
approximate treatment of the single particle mass operator with quartet 
condensate should be quite safe. With our separable ansatz for the two 
body interaction, we then arrive to expression 
(\ref{eq-quartetmassoperator}) of the mass operator.

\section{\label{sec-appendix-Immass}Preparation of mass operator for numerical application}

In this section, we describe the final expression for 
numerical calculation of the mass operator of Eq.~(\ref{eq-appqmo}).

We again give the approximated mass operator
\begin{widetext}
\begin{eqnarray}
&&
M^{\rm quartet}(k_1,\omega)
\nonumber \\
&=&\int \frac{d^3k_{2}}{(2\pi)^3} \frac{d^3k_{3}}{(2\pi)^3}
      \frac{d^3k_{4}}{(2\pi)^3}
  \frac{d^3k_{1'}}{(2\pi)^3} \frac{d^3k_{2'}}{(2\pi)^3} 
  \frac{d^3k_{1''}}{(2\pi)^3} \frac{d^3k_{2''}}{(2\pi)^3}
\nonumber \\
&\times&
e^{-2(\vec k_1-\vec k_2)^2/(4b^2)}
e^{-(\vec k_{1'}-\vec k_{2'})^2/(4b^2)}
e^{-(\vec k_{1''}-\vec k_{2'''})^2/(4b^2)}
(2\pi)^3\delta(\vec k_1+\vec k_2-\vec k_{1'}-\vec k_{2'})
\nonumber \\
&\times&
\varphi(|\vec k_{1'}|)\varphi(|\vec k_{2'}|)
\varphi(|\vec k_{3}|)\varphi(|\vec k_{4}|)
(2\pi)^3\delta(\vec k_{1'}+\vec k_{2'}+\vec k_{3}+\vec k_{4})
\nonumber \\
&\times&
\varphi(|\vec k_{1''}|)\varphi(|\vec k_{2''}|)
\varphi(|\vec k_{3}|)\varphi(|\vec k_{4}|)
(2\pi)^3\delta(\vec k_{1''}+\vec k_{2''}+\vec k_{3}+\vec k_{4}) 
\nonumber \\
&\times&
\frac{\bar f(k_{2}) \bar f(k_{3}) \bar f(k_{4})
     +f(k_{2})f(k_{3})f(k_{4})}
     {\omega+\varepsilon_{\vec k_{2}}
     +\varepsilon_{\vec k_{3}}+\varepsilon_{\vec k_{4}}}
\nonumber \\
&=&
\int \frac{d^3k_{2}}{(2\pi)^3} \frac{d^3K}{(2\pi)^3}
      \frac{d^3k}{(2\pi)^3}
  \frac{d^3K'}{(2\pi)^3} \frac{d^3k'}{(2\pi)^3} 
  \frac{d^3K''}{(2\pi)^3} \frac{d^3k''}{(2\pi)^3}
\nonumber \\
&\times&
e^{-2(\vec k_1-\vec k_2)^2/(4b^2)}
e^{-k'^2/b^2}e^{-k''^2/b^2}
(2\pi)^3\delta(\vec k_1+\vec k_2-\vec K')
\nonumber \\
&\times&
\varphi(|\frac{\vec K'}{2}+\vec k'|)
\varphi(|\frac{\vec K'}{2}-\vec k'|)
\varphi(|\frac{\vec K}{2}+\vec k|)
\varphi(|\frac{\vec K}{2}-\vec k|)
(2\pi)^3\delta(\vec K'+\vec K)
\nonumber \\
&\times&
\varphi(|\frac{\vec K''}{2}+\vec k''|)
\varphi(|\frac{\vec K''}{2}-\vec k''|)
\varphi(|\frac{\vec K}{2}+\vec k|)
\varphi(|\frac{\vec K}{2}-\vec k|)
(2\pi)^3\delta(\vec K''+\vec K) 
\nonumber \\
&\times&
\frac{\bar f(k_2) \bar f(|\frac{\vec K}{2}+\vec k|) 
      \bar f(|\frac{\vec K}{2}-\vec k|)
     +f(k_2)f(|\frac{\vec K}{2}+\vec k|)f(|\frac{\vec K}{2}-\vec k|)}
     {\omega+\varepsilon_{\vec k_{2}}
     +\varepsilon_{\frac{\vec K}{2}+\vec k}+\varepsilon_{\frac{\vec K}{2}-\vec k}}
\nonumber \\
&=&
\int \frac{d^3K}{(2\pi)^3}\frac{d^3k}{(2\pi)^3}
e^{-2(\vec k_1+\frac{\vec K}{2})^2/b^2}
\Bigl[\varphi(|\frac{\vec K}{2}+\vec k|)
\varphi(|\frac{\vec K}{2}-\vec k|)\Bigr]^2
\nonumber \\
&\times&
\frac{\bar f(|\vec k_1+\vec K|) 
      \bar f(|\frac{\vec K}{2}+\vec k|) 
      \bar f(|\frac{\vec K}{2}-\vec k|)
     +f(|\vec k_1+\vec K|)
      f(|\frac{\vec K}{2}+\vec k|)
      f(|\frac{\vec K}{2}-\vec k|)}
     {\omega+\frac{k_1^2}{2m}+\frac{\vec k_1\cdot \vec K}{m}
     +\frac{3K^2}{4m}+\frac{k^2}{m}-3\mu}
\nonumber \\
&\times&
\left[\int
\frac{d^3k'}{(2\pi)^3}
e^{-k'^2/b^2}
\varphi(|-\frac{\vec K}{2}+\vec k'|)\varphi(|-\frac{\vec K}{2}-\vec k'|)
\right]^2.
\end{eqnarray}
\end{widetext}

The imaginary part of the above mass operator
$-{\rm Im}M^{\rm quartet}(k_1,\omega+i\eta)$ is

\begin{eqnarray}
&&
-{\rm Im}M^{\rm quartet}(k_1,\omega+i\eta)
\nonumber \\
&=&
\pi \int \frac{d^3K}{(2\pi)^3}\frac{d^3k}{(2\pi)^3}
e^{-2(\vec k_1+\frac{\vec K}{2})^2/b^2}
\nonumber \\
&\times&
\Bigl[\varphi(|\frac{\vec K}{2}+\vec k|)
\varphi(|\frac{\vec K}{2}-\vec k|)\Bigr]^2
\nonumber \\
&\times&
\Bigl[\bar f(|\vec k_1+\vec K|) 
      \bar f(|\frac{\vec K}{2}+\vec k|) 
      \bar f(|\frac{\vec K}{2}-\vec k|)
\nonumber \\
&&
     +f(|\vec k_1+\vec K|)
      f(|\frac{\vec K}{2}+\vec k|)
      f(|\frac{\vec K}{2}-\vec k|)
\Bigr]
\nonumber \\
&\times&
\delta(\omega+\frac{k_1^2}{2m}+\frac{\vec k_1\cdot \vec K}{m}
     +\frac{3K^2}{4m}+\frac{k^2}{m}-3\mu)
\nonumber \\
&\times&
\Bigl[\int
\frac{d^3k'}{(2\pi)^3}
e^{-k'^2/b^2}
\varphi(|-\frac{\vec K}{2}+\vec k'|)
\varphi(|-\frac{\vec K}{2}-\vec k'|)
\Bigr]^2
\nonumber \\
\end{eqnarray}

\subsection{\label{subsec-k1zero}The $\vec k_1=0$ case}
The imaginary part of the mass operator for $\vec k_1=0$
is given by

\begin{eqnarray}
&&
-{\rm Im}M^{\rm quartet}(k_1=0,\omega+i\eta)
\nonumber \\
&=&
\pi \int \frac{d^3K}{(2\pi)^3}\frac{d^3k}{(2\pi)^3}
e^{-K^2/(4b^2)}
\nonumber \\
&\times&
\Bigl[\varphi(|\frac{\vec K}{2}+\vec k|)
\varphi(|\frac{\vec K}{2}-\vec k|)\Bigr]^2
\nonumber \\
&\times&
\Bigl[\bar f(K) 
      \bar f(|\frac{\vec K}{2}+\vec k|) 
      \bar f(|\frac{\vec K}{2}-\vec k|)
\nonumber \\
&&
     +f(K)
      f(|\frac{\vec K}{2}+\vec k|)
      f(|\frac{\vec K}{2}-\vec k|)
\Bigr]
\nonumber \\
&\times&
\delta(\omega+\frac{3K^2}{4m}+\frac{k^2}{m}-3\mu)
\nonumber \\
&\times&
\Bigl[\int
\frac{d^3k'}{(2\pi)^3}
e^{-k'^2/b^2}
\varphi(|-\frac{\vec K}{2}+\vec k'|)
\varphi(|-\frac{\vec K}{2}-\vec k'|)
\Bigr]^2
\nonumber \\
&=&
\pi \frac{2}{(2\pi)^4}\int_0^\infty dKK^2 \int_0^\infty dkk^2 
\int_{-1}^1dt  
e^{-K^2/(4b^2)}
\nonumber \\
&\times&
\Bigl[\varphi(|\frac{\vec K}{2}+\vec k|)
\varphi(|\frac{\vec K}{2}-\vec k|)\Bigr]^2
\nonumber \\
&\times&
\Bigl[\bar f(K) 
      \bar f(|\frac{\vec K}{2}+\vec k|) 
      \bar f(|\frac{\vec K}{2}-\vec k|)
\nonumber \\
&&
     +f(K)f(|\frac{\vec K}{2}+\vec k|)
      f(|\frac{\vec K}{2}-\vec k|)
\Bigr]
\nonumber \\
&\times&
\frac{m}{2k}\delta(k-\sqrt{3m\mu-m\omega-3K^2/4})
\nonumber \\
&\times&
\Bigl[\frac{1}{(2\pi)^2}
\int_0^\infty dk'k'^2 \int_{-1}^1dt'
e^{-k'^2/b^2}
\nonumber \\
&\times&
\varphi(|-\frac{\vec K}{2}+\vec k'|)
\varphi(|-\frac{\vec K}{2}-\vec k'|)
\Bigr]^2
\nonumber \\
&=&
\frac{2\pi m}{2(2\pi)^4}\int_0^{P_{max0}} dKK^2 p 
\int_{-1}^1dt  
e^{-K^2/(4b^2)}
\nonumber \\
&\times&
\Bigl[\varphi(|\frac{\vec K}{2}+\vec p|)
\varphi(|\frac{\vec K}{2}-\vec p|)\Bigr]^2
\nonumber \\
&\times&
\Bigl[\bar f(K) 
      \bar f(|\frac{\vec K}{2}+\vec p|) 
      \bar f(|\frac{\vec K}{2}-\vec p|)
\nonumber \\
&&
     +f(K)f(|\frac{\vec K}{2}+\vec p|)
      f(|\frac{\vec K}{2}-\vec p|)
\Bigr]
\nonumber \\
&\times&
\Bigl[\frac{1}{(2\pi)^2}
\int_0^\infty dk'k'^2 \int_{-1}^1dt'
e^{-k'^2/b^2}
\nonumber \\
&\times&
\varphi(|-\frac{\vec K}{2}+\vec k'|)
\varphi(|-\frac{\vec K}{2}-\vec k'|)
\Bigr]^2,
\end{eqnarray}
Here, in the last equation,
\begin{eqnarray}
p
&=&
\sqrt{3m\mu-m\omega-\frac{3K^2}{4}},
\\
P_{max0}
&=&
\sqrt{\frac{4}{3}(3m\mu-m\omega)},
\\
t
&=&
(\vec p\cdot \vec K)/(pK),
\\
t'
&=&
(\vec k'\cdot \vec K)/(k'K).
\end{eqnarray}

\subsection{\label{subsec-k1nonzero}The $\vec k_1 \neq 0$ case}
On the other hand, 
$-{\rm Im}M^{\rm quartet}$ for $\vec k_1 \neq 0$
is represented by

\begin{eqnarray}
&&
-{\rm Im}M^{\rm quartet}(k_1,\omega+i\eta)
\nonumber \\
&=&
\frac{\pi}{(2\pi)^4}\int_0^{\infty} dKK^2 
\int^{1}_{-1}dt_1 \int_0^\infty dk k^2 \int^{1}_{-1} dt 
\nonumber \\
&\times&
e^{-2(k_1^2+K^2/4+k_1Kt_1)/b^2}
\Bigl[\varphi(|\frac{\vec K}{2}+\vec k|)
\varphi(|\frac{\vec K}{2}-\vec k|)\Bigr]^2
\nonumber \\
&\times&
\Bigl[\bar f(|\vec k_1+\vec K|) 
      \bar f(|\frac{\vec K}{2}+\vec k|) 
      \bar f(|\frac{\vec K}{2}-\vec k|)
\nonumber \\
&&
     +f(|\vec k_1+\vec K|)f(|\frac{\vec K}{2}+\vec k|)
      f(|\frac{\vec K}{2}-\vec k|)
\Bigr]
\nonumber \\
&\times&
\delta(\omega+\frac{k_1^2}{2m}+\frac{k_1Kt_1}{m}
     +\frac{3K^2}{4m}+\frac{k^2}{m}-3\mu)
\nonumber \\
&\times&
\Bigl[\frac{1}{(2\pi)^2}
\int dk'k'^2 \int^1_{-1}dt' 
e^{-k'^2/b^2}
\nonumber \\
&\times&
\varphi(|-\frac{\vec K}{2}+\vec k'|)
\varphi(|-\frac{\vec K}{2}-\vec k'|)
\Bigr]^2
\nonumber \\
&=&
\frac{\pi}{(2\pi)^4}\int_0^{\infty} dKK^2 
\int^{1}_{-1}dt_1 \int_0^\infty dk k^2 \int^{1}_{-1} dt 
\nonumber \\
&\times&
e^{-2(k_1^2+K^2/4+k_1Kt_1)/b^2}
\Bigl[\varphi(|\frac{\vec K}{2}+\vec k|)
\varphi(|\frac{\vec K}{2}-\vec k|)\Bigr]^2
\nonumber \\
&\times&
\Bigl[\bar f(|\vec k_1+\vec K|) 
      \bar f(|\frac{\vec K}{2}+\vec k|) 
      \bar f(|\frac{\vec K}{2}-\vec k|)
\nonumber \\
&&
     +f(|\vec k_1+\vec K|)
      f(|\frac{\vec K}{2}+\vec k|)
      f(|\frac{\vec K}{2}-\vec k|)
\Bigr]
\nonumber \\
&\times&
\frac{m}{k_1K}\delta(t_1-\frac{3m\mu-m\omega-\frac{k_1^2}{2}
     -\frac{3K^2}{4}-k^2}{k_1K})
\nonumber \\
&\times&
\Bigl[\frac{1}{(2\pi)^2}
\int dk'k'^2 \int^1_{-1}dt' 
e^{-k'^2/b^2}
\nonumber \\
&\times&
\varphi(|-\frac{\vec K}{2}+\vec k'|)
\varphi(|-\frac{\vec K}{2}-\vec k'|)
\Bigr]^2.
\end{eqnarray}
Here, the following condition has to be satisfied:
\begin{eqnarray}
-1< \frac{3m\mu-m\omega-\frac{k_1^2}{2}
     -\frac{3K^2}{4}-k^2}{k_1K} <1.
\end{eqnarray}
Therefore
\begin{eqnarray}
k^2> 3m\mu-m\omega-\frac{k_1^2}{2}
     -\frac{3K^2}{4}-k_1K>0 
\end{eqnarray}
and
\begin{eqnarray}
0<k^2< 3m\mu-m\omega-\frac{k_1^2}{2}
     -\frac{3K^2}{4}+k_1K.
\label{eq-limitkint}
\end{eqnarray}
From the above equation, we obtain the limits of integration 
with respect to $k$:
\begin{eqnarray}
&&
p_{min}(K)=\sqrt{\max(3m\mu-m\omega-\frac{k_1^2}{2}
     -\frac{3K^2}{4}-k_1K,0)}
\nonumber \\
&&
<k<\sqrt{3m\mu-m\omega-\frac{k_1^2}{2}
     -\frac{3K^2}{4}+k_1K}=p_{max}(K)
\nonumber \\
\end{eqnarray}
Besides, from Eq.~(\ref{eq-limitkint}),
\begin{eqnarray}
3m\mu-m\omega-\frac{k_1^2}{2}
     -\frac{3K^2}{4}+k_1K >0
\end{eqnarray}
has to be satisfied. 
Therefore
\begin{eqnarray}
&&
P_{min}=\max\biggl[\frac{2}{3}\Bigl(k_1-
\sqrt{9m\mu -3m\omega -\frac{k_1^2}{2}}\Bigr),0\biggr]
\nonumber \\
&&
<K<
\frac{2}{3}\left(k_1+
\sqrt{9m\mu -3m\omega -\frac{k_1^2}{2}}\right)=P_{max}
\nonumber \\
\end{eqnarray}

\subsection{Summary of integrals in Im$M^{\rm quartet}$}

With the Appendices \ref{subsec-k1zero} and \ref{subsec-k1nonzero},
we can express $-{\rm Im}M^{\rm quartet}(k_1,\omega+i\eta)$ by:

\vspace{10mm}

\noindent If $9m\mu -3m\omega -\frac{k_1^2}{2}<0$,
\begin{eqnarray}
-{\rm Im}M^{\rm quartet}(k_1,\omega+i\eta)=0.
\end{eqnarray}

\noindent Else:

If $k_1=0$,
\begin{eqnarray}
&&
-{\rm Im}M^{\rm quartet}(k_1,\omega+i\eta)
\nonumber \\
&=&
\frac{2\pi m}{2(2\pi)^4}\int_0^{P_{max0}} dKK^2 p e^{-K^2/(4b^2)}
\nonumber \\
&\times&
\int_{-1}^1dt  
\left[\varphi(|\frac{\vec K}{2}+\vec p|)
\varphi(|\frac{\vec K}{2}-\vec p|)\right]^2
\nonumber \\
&\times&
\Bigl[\bar f(K) 
      \bar f(|\frac{\vec K}{2}+\vec p|) 
      \bar f(|\frac{\vec K}{2}-\vec p|)
\nonumber \\
&&
     +f(K)f(|\frac{\vec K}{2}+\vec p|)
       f(|\frac{\vec K}{2}-\vec p|)
\Bigr]
\nonumber \\
&\times&
\Bigl[\frac{1}{(2\pi)^2}
\int_0^\infty dk'k'^2 e^{-k'^2/b^2}
\int_{-1}^1dt' 
\nonumber \\
&\times&
\varphi(|-\frac{\vec K}{2}+\vec k'|)
\varphi(|-\frac{\vec K}{2}-\vec k'|)
\Bigr]^2, 
\nonumber \\
&&
p=\sqrt{3m\mu-m\omega-\frac{3K^2}{4}},
\nonumber \\
&&
P_{max0}=\sqrt{\frac{4}{3}(3m\mu-m\omega)},
\nonumber \\
&&t=(\vec k\cdot \vec K)/(kK),
\qquad t'=(\vec k'\cdot \vec K)/(k'K),
\end{eqnarray}

while if $k_1\neq0$,

\begin{eqnarray}
&&
-{\rm Im}M^{\rm quartet}(k_1,\omega+i\eta)
\nonumber \\
&=&
\frac{\pi m}{(2\pi)^4k_1}\int_{P_{min}}^{P_{max}} dKK
\int_{p_{min}}^{p_{max}} dk k^2 
\nonumber \\
&\times&
e^{-2(3m\mu-m\omega+k_1^2/2-K^2/2-k^2)/b^2}
\nonumber \\
&\times&
\int^{1}_{-1} dt 
\Bigl[\varphi(|\frac{\vec K}{2}+\vec k|)
\varphi(|\frac{\vec K}{2}-\vec k|)\Bigr]^2
\nonumber \\
&\times&
\Bigl[\bar f(\sqrt{6m\mu-2m\omega-\frac{K^2}{2}-2k^2}) 
      \bar f(|\frac{\vec K}{2}+\vec k|) 
      \bar f(|\frac{\vec K}{2}-\vec k|)
\nonumber \\
&&
+f(\sqrt{6m\mu-2m\omega-\frac{K^2}{2}-2k^2})
     f(|\frac{\vec K}{2}+\vec k|)f(|\frac{\vec K}{2}-\vec k|)
\Bigr]
\nonumber \\
&\times&
\Bigl[\frac{1}{(2\pi)^2}
\int_0^\infty dk'k'^2 e^{-k'^2/b^2}
\int^1_{-1}dt' 
\nonumber \\
&\times&
\varphi(|-\frac{\vec K}{2}+\vec k'|)
\varphi(|-\frac{\vec K}{2}-\vec k'|)
\Bigr]^2 ,
\nonumber \\
&&
p_{min}(K)=\sqrt{\max[3m\mu-m\omega-\frac{k_1^2}{2}
     -\frac{3K^2}{4}-k_1K,0]},
\nonumber \\
&&
p_{max}(K)=\sqrt{3m\mu-m\omega-\frac{k_1^2}{2}
     -\frac{3K^2}{4}+k_1K}, 
\nonumber \\
&&
P_{min}=\max\left[\frac{2}{3}\left(k_1-
\sqrt{9m\mu -3m\omega -\frac{k_1^2}{2}}\right),0\right], 
\nonumber \\
&&
P_{max}=
\frac{2}{3}\left(k_1+
\sqrt{9m\mu -3m\omega -\frac{k_1^2}{2}}\right),
\nonumber \\
&&
\nonumber \\
&&t=(\vec k\cdot \vec K)/(kK),
\qquad t'=(\vec k'\cdot \vec K)/(k'K).
\end{eqnarray}

Re$M^{\rm quartet}$ is calculated by Eq.~(\ref{eq-remassoperator}).




\begin{thebibliography}{99}

\bibitem{alphaBECfinite}
A. Tohsaki, H. Horiuchi, P. Schuck, and G. R\"opke,
Phys. Rev. Lett. {\bf 87}, 192501 (2001);
Y. Funaki, T. Yamada, H. Horiuchi, G. R\"opke, P. Schuck, and A. Tohsaki,
Phys. Rev. Lett. {\bf 101}, 082502 (2008).



\bibitem{trk10}
S. Typel1, G. R\"opke, T. Kl\"ahn, D. Blaschke, and H. H. Wolter,
Phys. Rev. C {\bf 81}, 015803 (2010).


\bibitem{wlo09}
A. N. Wenz, T. Lompe, T. B. Ottenstein, F. Serwane, G. Z\"orn, and S. Jochim,
Phys. Rev. A {\bf 80}, 040702(R) (2009).


\bibitem{whh09}
J. R. Williams, E. L. Hazlett, J. H. Huckans, 
R. W. Stites, Y. Zhang, and K. M. O'Hara,
Phys. Rev. Lett. {\bf 103}, 130404 (2009).






\bibitem{rzh07}
\'Akos Rapp, Gergely Zar\'and, Carsten Honerkamp, and Walter Hofstetter,
Phys. Rev. Lett. {\bf 98}, 160405 (2007).

\bibitem{gbl08}
X. W. Guan, M. T. Batchelor, C. Lee, and H.-Q. Zhou
Phys. Rev. Lett. {\bf 100}, 200401 (2008). 

\bibitem{kdd09}
A. Kantian, M. Dalmonte, S. Diehl, W. Hofstetter, P. Zoller, and A. J. Daley,
Phys. Rev. Lett. {\bf 103}, 240401 (2009). 

\bibitem{fsw09}
S. Floerchinger, R. Schmidt, and C. Wetterich,
Phys. Rev. A {\bf 79}, 053633 (2009). 

\bibitem{edo09}
Beatriz Errea, Jorge Dukelsky, and Gerardo Ortiz,
Phys. Rev. A {\bf 79}, 051603 (2009).

\bibitem{silva09}
Theja N. De Silva,
Phys. Rev. A {\bf 80}, 013620 (2009). 

\bibitem{mds09}
Rafael A. Molina, Jorge Dukelsky, and Peter Schmitteckert,
Phys. Rev. A {\bf 80}, 013616 (2009). 

\bibitem{acl09}
P. Azaria, S. Capponi, and P. Lecheminant,
Phys. Rev. A {\bf 80}, 041604 (2009).

\bibitem{mis10}
Shin-ya Miyatake, Kensuke Inaba, and Sei-ichiro Suga,
Phys. Rev. A {\bf 81}, 021603 (2010). 







\bibitem{km09}
H. Kamei and K. Miyake,
J. Phys. Soc. Jpn. {\bf 74}, 1911 (2005).

\bibitem{sg99}
A. S. Stepanenko and J. M. Gunn,
arXiv:cond-mat/9901317.

\bibitem{schlottmann94}
P. Schlottmann, J. Phys. Condens. Matter {\bf 6}, 1359 (1994).

\bibitem{wu05}
C. Wu,
Phys. Rev. Lett. {\bf 95}, 266404 (2005).

\bibitem{rcl09}
G. Roux, S. Capponi, P. Lecheminant, and P. Azaria,
Eur. Phys. J. B {\bf 68}, 293 (2009).



\bibitem{hn}
M.~Hasuo and N.~Nagasawa, 
in {\it Bose-Einstein Condensation} edited by
A. Greffin, D. W. Snoke and S. Stringari
(Cambridge University Press, Cambridge, 1995), p. 487.

\bibitem{ms}
S. A. Moskalenko and D. W. Snoke, 
{\it Bose-Einstein Condensation of Excitons and Biexcitons} 
(Cambridge University Press, Cambridge, 2000).



\bibitem{slr09}
T. Sogo, R. Lazauskas, G. R\"opke, and P. Schuck,
Phys. Rev. C {\bf 79}, 051301 (2009).



\bibitem{rss98}
G. R\"opke, A. Schnell, P. Schuck, and P. Nozi\`eres,
Phys. Rev. Lett. {\bf 80}, 3177 (1998).




\bibitem{thouless}
D. J. Thouless, Ann. Phys. {\bf 10}, 553 (1960).

\bibitem{agd}
A. A. Abrikosov, L. P. Gorkov, and I. E. Dzyaloshinski,
{\it Mothods of quantum field theory in statistical physics},
(Dover, New York, 1975).


\bibitem{fw}
A. L. Fetter and J. D. Walecka,
{\it Quantum Theory of Many-Particle Systems},
(Dover, New York, 2003).


\bibitem{rs}P. Ring and P. Schuck,
{\it The Nuclear Many-Body Problem},
(Springer-Verlag, New York, 1980)


\bibitem{hmd02}
J.G. Hirsch, A. Mariano, J. Dukelsky, and P. Schuck, 
Ann. Phys. {\bf 296}, 187 (2002).


\bibitem{bhh86}
A. H. Blin, R. W. Hasse, B. Hiller, P. Schuck, and C. Yannouleas,
Nucl. Phys. {\bf A456}, 109 (1986).



\bibitem{as89}
S. Adachi and P. Schuck, 
Nucl. Phys. {\bf A496}, 485 (1989).



\bibitem{drs98}
J. Dukelsky, G. R\"opke, and P. Schuck,
Nucl. Phys. {\bf A628}, 17 (1998).


\bibitem{mahan}
G. D. Mahan,
{\it Many-Particle Physics},
(Plenum, New York, 1990).

\bibitem{ns85}
P. Nozi\`eres and S. Schmitt-Rink,
J. Low Temp. Phys. {\bf 59}, 195 (1985).


\bibitem{bbb83}
G. F. Bertsch, P. F. Bortignon, and R. A. Broglia,
Rev.~Mod.~Phys. {\bf 55}, 287 (1983).


\bibitem{aih83}
K. Ando, A. Ikeda and, G. Holzwarth,
Z.~Phys. {\bf A310} 223 (1983).


\bibitem{rms}
G. R\"opke, L. M\"unchow, and H. Schulz,
Nucl. Phys. {\bf A379}, 536 (1982);
G. R\"opke, M. Schmidt, L. M\"unchow, and H. Schulz,
Nucl. Phys. {\bf A399}, 587 (1983).



\end{thebibliography}
\end{document}